\RequirePackage{fix-cm}
\documentclass[smallextended]{svjour3}       
\smartqed  
\usepackage[pdftex]{graphicx}
\usepackage{multirow}
\newcommand\Rey{\mbox{\textrm{Re}}}            
\newcommand\Tay{\mbox{\textrm{Ta}}}            
\newcommand\Ros{\mbox{\textrm{Ro}}}            
\newcommand\uin{u_{\rm in}}
\newcommand{\eref}[1]{Eq.~(\ref{#1})}
\newcommand{\fref}[1]{Fig.~\ref{#1}}
\newcommand{\Fref}[1]{Figure~\ref{#1}}
\newcommand{\tref}[1]{Table~\ref{#1}}

\journalname{Int J Adv Eng Sci Appl Math}
\begin{document}

\title{DNS of Taylor--Couette flow between counter-rotating cylinders\\ at small radius ratio}

\titlerunning{Taylor--Couette flow at small radius ratio}        

\author{Ryo Tanaka \and Takuya Kawata \and Takahiro Tsukahara}

\institute{Ryo Tanaka 
       \at 
           \email{7517640@ed.tus.ac.jp}
           \and
           Takuya Kawata
       \at 
           \email{kawata@rs.tus.ac.jp}
           \and
           Takahiro Tsukahara
       \at Department of Mechanical Engineering, Tokyo University of Science, \\
           2641 Yamazaki, Noda-shi, Chiba, 278-5810 Japan\\
           Tel.: +81-4-7122-9352\\
           Fax: +81-4-7123-9814\\
           \email{tsuka@rs.tus.ac.jp}
}

\date{}

\maketitle

\begin{abstract}
A counter-rotating Taylor--Couette flow with relatively small radius ratios of $\eta = 0.2$--0.5 was investigated over a wide range of the Reynolds number, from laminar to turbulent regime, by means of three-dimensional direct numerical simulations. 
We investigated the $\eta$ dependence of the flow structure and determined a critical value between $\eta=0.2$ and 0.3, below which, the stable outer cylinder side exhibited a modal structure that was different from the Taylor-vortex flow on the inner side. 
At $\eta \geq 0.3$, the Taylor-vortex on the unstable inner side dominated the entire flow field between the cylinders, whose footprints were observed in the vicinity of the outer cylinder wall. 
However, for $\eta=0.2$, the influence from the inner side was limited up to the centre of the cylinder gap. Moreover, on the stable outer cylinder side, there appeared a modal structure that was axially homogeneous, azimuthally periodic, and similar to the Tollmien-Schlichting (TS) instability wave. 
As the Reynolds number increased with a fixed $\eta=0.2$, the modal structure changed its azimuthal wavenumber and thickened radially in the wall unit. 
Although the Reynolds shear stress on the outer side remained approximately zero, the intensity of the velocity fluctuations was comparable to the Taylor-vortex flows in the central part.

\keywords{DNS \and Flow instability \and Taylor--Couette flow \and Transition \and Wall turbulence}
\end{abstract}

\section{Introduction}
\label{intro}

A flow between two differentially rotating coaxial cylinders, i.e., the Taylor--Couette flow, provides a canonical system to analyse centrifugal fluid instabilities. 
This flow depends on the cylinder rotational velocities and directions, and has been investigated under various conditions. 
The laminar-turbulence transition mechanism is determined by the direction of the cylinder relative rotation. 
In cases of co-rotation or only inner cylinder rotation, the flow undergoes a supercritical transition owing to a linear centrifugal instability, where the Taylor-vortices appear by exceeding the well-defined critical Reynolds number. 
In the case of counter-rotation, the transition will be subcritical, the flow will become turbulent through bypass transition, and the coexistence of laminar and turbulent regions, such as the spiral turbulence and turbulent spot, may appear. 
Relevant studies have been reported widely and reviewed comprehensively \cite[and reference therein]{Tagg94,Fardin14,Grossmann15}. 

Many studies on the Taylor--Couette flow have been conducted for various Reynolds numbers and radius ratios of $\eta \geq 0.5$.
Here, the following radius ratio
\begin{equation}
\eta = \frac{r_{\rm in}}{r_{\rm out}}
\end{equation}
is defined by the two radii (denoted as $r_{\rm in}$ and $r_{\rm out}$, respectively) of the inner and outer cylinders.
Coles~\cite{RefA} investigated fluid motion with an increasing rotational velocity of the inner cylinder and observed the behaviour of the Taylor-vortex and spiral turbulence for the limited $\eta$ of $\approx 0.88$. 
He addressed the hysteresis of the state transitions in terms of the number of Taylor-cells, and also observed the spiral band of the turbulence as a result of subcritical transition, owing to an outer cylinder rotation that was much faster than the inner one. 
Andereck et al. \cite{RefB} have provided a map of the flow-state transition for different cylinder rotations at a radius ratio of $\eta = 0.883$, which has been an indispensable guide for experimental and theoretical investigation. Goharzadeh \& Mutabazi~\cite{RefC} reported that the turbulent spots grow in size as the inner cylinder rotational velocity increases, and eventually form a spiral turbulence. Dong~\cite{Dong08} investigated the counter-rotating case for different Reynolds numbers at the medium radius ratio of $\eta=0.5$ and reported the feature of the Taylor--Couette turbulence. 
Dong and Zheng~\cite{Dong11} investigated the behaviour of the spiral turbulence in the case where the rotational rate of the outer cylinder was fixed and only the inner rotational rate was changed for $\eta = 0.89$. They found that the transition from spiral turbulence to featureless turbulence progressed from the inner cylinder toward the outer one. 
Ostilla-Monico et al.~\cite{RefH} investigated the flow transition for the different radius ratios of $\eta = 0.909$, 0.714, and 0.5. 
They reported the same transition scenario at the same Taylor number for $\eta = 0.909$ and 0.714. They also reported that the critical Taylor number increased as the radius ratio decreased in the range of $\eta \leq 0.714$. 
Liscchke \& Roesner~\cite{RefF} experimentally investigated the spiral turbulence for different radius ratios, and decreased the Reynolds number with a higher radius ratio. 
The plane Couette flow with a spanwise system rotation corresponding to the Taylor--Couette flow with a limit of $\eta \to 1$ has been experimentally investigated by Tsukahara et al.~\cite{Tsuka10}, who carried out visualisation for a wide parameter range and provided a flow map, similar to the study by Andereck et al.~\cite{RefB}.
Although the moderate or large radius ratios of $\eta \geq 0.5$ have been investigated by several researchers, the flow structures for small radius ratios have not been clarified yet.
Moreover, the effect of the counter rotation, which may give rise to flow stabilization on the outer cylinder side, is also an open issue. 
A wide gap between the cylinders with a low $\eta$ would allow us to determine an interactive competition between the flow instability (on the inner side of the cylinder) and the stabilization effect (on the outer side), whose competition must be related to the complicated subcritical transition process of the counter-rotating Taylor--Couette flow.

In this study, we focused on the counter-rotating Taylor--Couette flow with low and medium radius ratios. We investigated the flow structure and turbulence statistics in the radius-ratio range of $0.2 \leq \eta \leq 0.5$ over a wide range of the Reynolds number, from the laminar to turbulent regime, by a three-dimensional direct numerical simulation (DNS).

\section{Numerical procedure}

\subsection{Flow system and governing equations}
\label{sec:system}

\begin{figure}[t]
\begin{center}
\includegraphics[width=0.65\hsize]{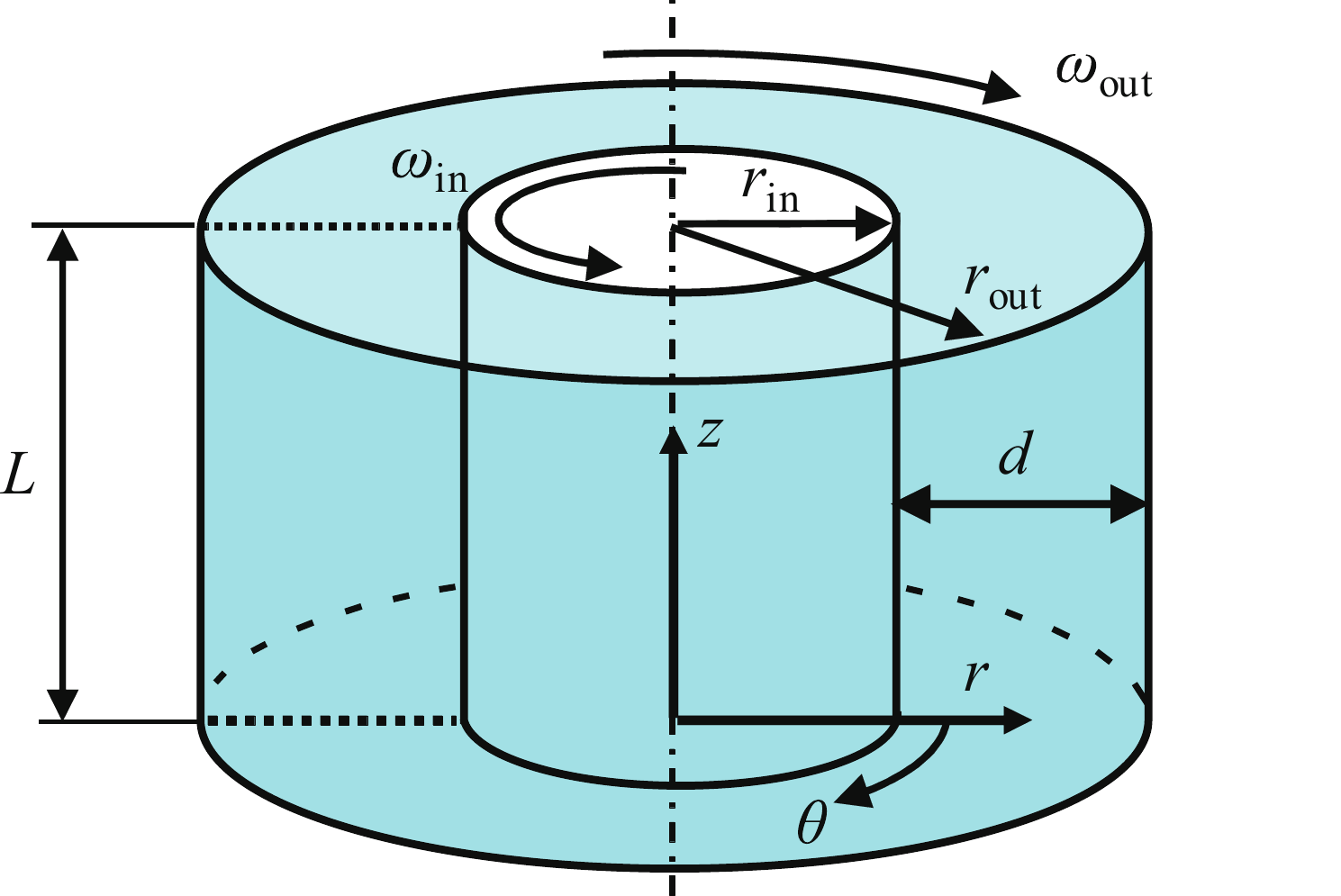}
\caption{Configuration of Taylor--Couette system consisting of two concentric cylinders. The domain of interest had an axial dimension $L$ and azimuthal dimension $2\pi$, with periodic boundaries in both the $z$ and $\theta$ directions. The inner cylinder with a radius $r_{\rm in}$ rotated at the angular velocity of $\omega_{\rm in}$, while the outer cylinder with a radius of $r_{\rm out}$ counter-rotated at the angular velocity of $\omega_{\rm out}$ $(<0< \omega_{\rm in})$.}
\label{fig:analysis}
\end{center}
\end{figure}

We considered the incompressible Newtonian flows between the two counter-rotating concentric cylinders, as shown in \fref{fig:analysis}. In the current simulations, the cylinders had an axial dimension of $L = 3.2d$, and the periodicity was assumed at both ends in the azimuthal direction: for instance, ${\bf u}(r,\theta,0)={\bf u}(r,\theta,L)$ for the velocity. This setting can be used to approximate two infinitely long cylinders. The wall-normal height from the inner cylinder surface was used throughout this paper instead of $r$, and is denoted as $r^*$, which is dimensionless based on the gap width $d=r_{\rm out}-r_{\rm in}$, as follows:
\begin{equation}
 r^* = \frac{r-r_{\rm in}}{r_{\rm out}-r_{\rm in}}.
\label{eq:r*}
\end{equation}
All of the lengths are normalized by the cylinder gap $d$, while the velocities are done by the inner cylinder surface velocity $\uin$. 
The inner cylinder rotated with a positive angular velocity $\omega_{\rm in} > 0$ around the $z$-axis, while the outer cylinder counter-rotated at $\omega_{\rm out} < 0$. 
This study has limitations in the case where: 
\begin{equation}
\uin \equiv r_{\rm in} \omega_{\rm in}=-r_{\rm out} \omega_{\rm out}.
\label{eq:uin}
\end{equation}
Hereafter, we will use a single Reynolds number as a control parameter with the kinematic viscosity $\nu$, which is uniquely defined as follows:
\begin{equation}
 \Rey = \frac{\uin d}{\nu}
\end{equation}
Note that the Reynolds number based on the outer cylinder surface velocity should be equal to $\Rey$.
Additionally, the Taylor number and Rossby number are defined based on the radii and rotational angular velocities of the cylinders, respectively,
as follows:
\begin{equation}
{\rm Ta} = \frac{1}{4}\left( \frac{r_{\rm in}+r_{\rm out}}{2\sqrt{r_{\rm in}r_{\rm out}}} \right)^4 
 \frac{\left(r_{\rm in} + r_{\rm out} \right)^2 \left(\omega_{\rm in} - \omega_{\rm out} \right)^2 d^2}{\nu^2},
\end{equation}
and
\begin{equation}
{\rm Ro} = \frac{r_{\rm in} | \omega_{\rm in}-\omega_{\rm out}|}{2\omega_{\rm out}d},
\end{equation}

These two non-dimensional parameters are of universal importance in determining the flow pattern regardless of the radius ratio: the former quantifies the importance of the centrifugal (or inertial) forces relative to the viscous force, while the latter indicates the ratio of the inertial force to the Coriolis force.
Under the current condition of \eref{eq:uin}, ${\rm Ta}$ is no longer independent from $\eta$ and $\Rey$, while ${\rm Ro}^{-1} = 2(1-\eta)/(1+\eta)$. 

We employed the cylindrical coordinates to describe the flow, where the $r$-, $\theta$- and $z$-axes are defined in the radial, azimuthal, and axial directions, respectively. 
The governing equations are the dimensionless continuity and Navier-Stokes equations:
\begin{eqnarray} 
&\displaystyle \frac{\partial u_x}{\partial x} + \frac{1}{r} \frac{\partial \left(ru_r\right)}{\partial r} + \frac{1}{r}\frac{\partial u_{\theta}}{\partial \theta}
= 0, \label{eq:continuity} & \\
&\displaystyle \frac{\partial u_r}{\partial t}+\left( \mathbf{u} \cdot \nabla \right)u_r-\frac{u_{\theta}^2}{r}
= -\frac{\partial p}{\partial r}+\frac{1}{\Rey} \left( \nabla^2u_r - \frac{u_r}{r^2} - \frac{2}{r^2}\frac{\partial u_{\theta}}{\partial \theta} \right),&
\label{eq:NSr} \\
&\displaystyle \frac{\partial u_{\theta}}{\partial t}+\left( \mathbf{u} \cdot \nabla \right)u_{\theta}+\frac{u_{r} u_\theta}{r}
=-\frac{1}{r}\frac{\partial p}{\partial \theta}+\frac{1}{\Rey}\left( \nabla^2u_{\theta} -\frac{u_\theta}{r^2}+\frac{2}{r^2}\frac{\partial u_{r}}{\partial \theta}\right), &
\label{eq:NStheta} \\
&\displaystyle \frac{\partial u_z}{\partial t}+\left(\mathbf{u} \cdot \nabla \right)u_z 
= -\frac{\partial p}{\partial z}+\frac{1}{\Rey} \nabla^2 u_z,&
\label{eq:NSx} 
\end{eqnarray}
where $(u_r, u_\theta,u_z)$ are the velocity components in the $(r,\theta,z)$ directions, $p$ the pressure, and $t$ the time. 
All quantities in the above equations are scaled by $d$ and/or $\uin$.

\subsection{Numerical details and code validation}
\label{sec:method}

Equations~(\ref{eq:continuity})-(\ref{eq:NSx}) were discretised by the finite difference method, as follows: for spatial discretisation, the fourth-order central difference scheme was adopted for the azimuthal and axial directions, while the second-order scheme was applied in the radial direction. 
The time advancement was carried out by the second-order Adams--Bashforth method, with the exception of the viscous term in the $r$-direction, for which the second-order Crank--Nicolson method was employed. 

\begin{table}
\caption{Computational condition for DNS of counter-rotating Taylor--Couette flow.}
\label{tab:numerical}
\begin{tabular}{lrlll}
\hline\noalign{\smallskip}
$\eta$         &$\Rey$~&$N_r\times N_\theta \times N_z$&$\Tay$   &$\Ros^{-1}$            \\ \noalign{\smallskip}\hline\noalign{\smallskip}
\multirow{2}*{0.5}& 500&$128\times256\times256$&$1.60\times 10^6$&\multirow{2}*{$-0.667$}\\ 
                  &4000&$128\times512\times256$&$1.02\times 10^8$&                       \\ \noalign{\smallskip}\hline\noalign{\smallskip}
\multirow{2}*{0.4}& 500&$128\times256\times256$&$2.25\times 10^6$&\multirow{2}*{$-0.857$}\\ 
                  &4000&$256\times512\times512$&$1.44\times 10^8$&                       \\ \noalign{\smallskip}\hline\noalign{\smallskip}
\multirow{2}*{0.3}& 500&$128\times256\times256$&$3.93\times 10^6$&\multirow{2}*{$-1.08$} \\ 
                  &4000&$256\times512\times512$&$2.52\times 10^8$&                       \\ \noalign{\smallskip}\hline\noalign{\smallskip}
\multirow{7}*{0.2}& 500&$128\times256\times256$&$1.04\times 10^7$&\multirow{6}*{$-1.33$} \\ 
                  &1000&$128\times256\times256$&$4.20\times 10^7$&                       \\ 
                  &1250&$128\times256\times256$&$6.56\times 10^7$&                       \\ 
                  &1375&$128\times256\times256$&$7.94\times 10^7$&                       \\ 
                  &1500&$128\times256\times256$&$9.45\times 10^7$&                       \\ 
                  &2000&$128\times256\times256$&$1.68\times 10^8$&                       \\ 
                  &4000&$256\times512\times256$&$6.72\times 10^8$&                       \\ 
\noalign{\smallskip}\hline
\end{tabular}
\end{table}

We conducted a series of simulations over the range of $0.2 \leq \eta \leq 0.5$ and $500 \leq \Rey \leq 4000$. 
The number of grids and the corresponding Taylor and Rossby numbers in each case are listed in Table~\ref{tab:numerical}. 
The computational domain size was $d \times 2\pi \times 3.2d$ in $(r, \theta, z)$: see \fref{fig:analysis}. 
Note again that the periodic boundary conditions were used in the $\theta$ and $z$ directions, while the non-slip condition was applied to the cylinder walls. 
The initial condition in each simulation was the laminar flow at the corresponding $\eta$. A fully-developed (statistically steady) flow field was achieved by a long-time simulation. 
The resulting friction velocities, the friction Reynolds numbers, and the spatial resolutions in the wall units are summarized in Table~\ref{tab:utau}. 
Note that the friction velocity $u_\tau$ was calculated from the mean velocity gradient on the surface of each cylinder. 

\begin{figure}
\includegraphics[height=38mm]{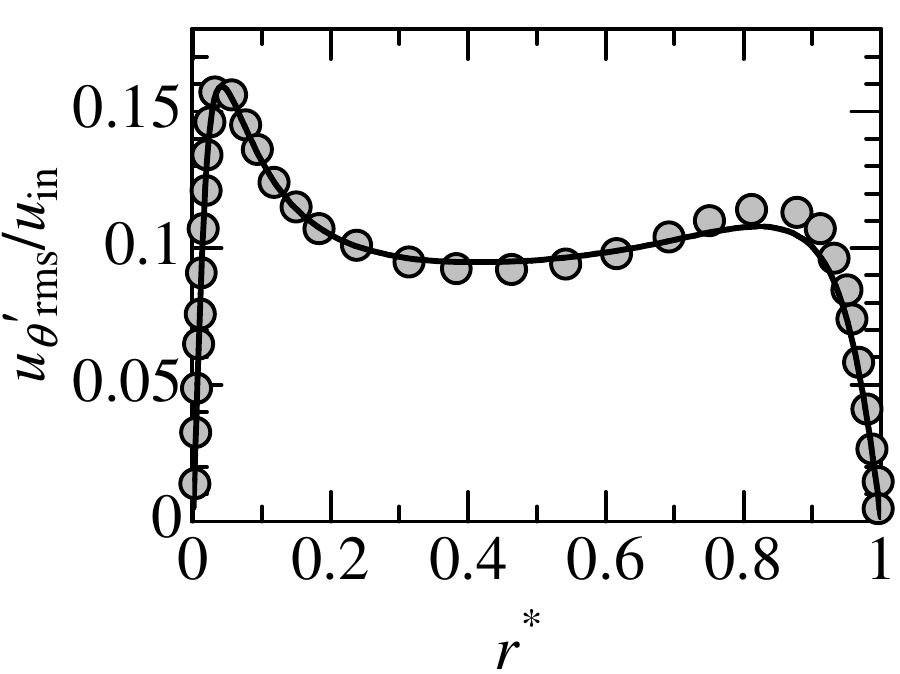} \hspace{0.5mm}
\includegraphics[height=38mm]{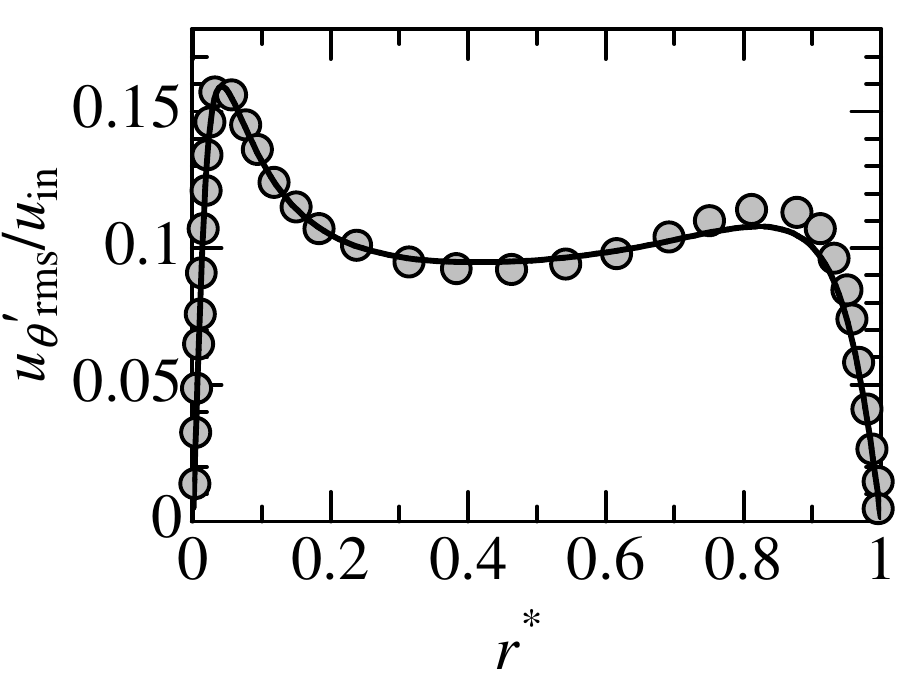} \hspace{0.5mm}
\includegraphics[height=38mm]{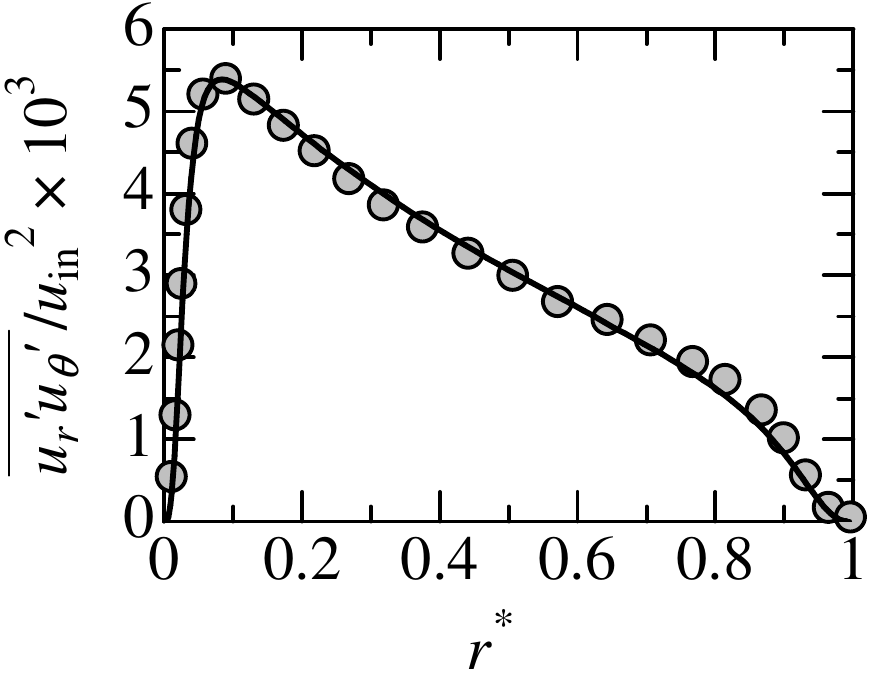}
\caption{Code validation by comparison with previous DNS study for $\eta = 0.5$ at $\Rey=4000$. Left: mean azimuthal velocity; middle: root-mean-square of azimuthal velocity fluctuation $u^\prime_{\theta, \mathrm{rms}}$; right: Reynolds shear stress $-\overline{u'_r u'_\theta}$. Lines and symbols represent results obtained by this study with the finite-difference method and results by Dong et al.~\cite{Dong08} obtained with the DNS of the spectral method, respectively.}
\label{fig:previous}
\end{figure}

To validate the present numerical simulation, the profiles of the mean azimuthal velocity $\overline{u_\theta}$, the azimuthal velocity fluctuation $u_{\theta, \mathrm{rms}}^\prime$, and the Reynolds shear stress $\overline{u_r^\prime u_\theta^\prime}$ at $\eta=0.5$ and $\Rey=4000$ were compared with those in existing DNS data~\cite{Dong08}. 
A mean value represented by an overbar was quantity ensemble-averaged in $\theta$, $z$, and $t$, after achieving a statistically steady state. 
As shown in \fref{fig:previous}, the numerical results obtained by this study are in good quantitative agreement with the reference study.

\begin{table}
\caption{Obtained friction velocity and grid resolution in wall unit: $\Delta y_{\rm in}$ represents the first velocity-vector height from the inner cylinder surface; $r\Delta \theta$ the azimuthal grid resolution (here, $\Delta \theta = 2\pi/N_\theta$), and $\Delta z_{\rm max}^+$ the axial grid resolution normalized by $\nu/u_{\tau,{\rm in}}$ on the inner cylinder surface, which is $1/\eta$-times larger than that of the outer cylinder. The friction Reynolds number $\Rey_\tau$ is based on $d/2\nu$ and either on $u_{\tau,{\rm in}}$ or $u_{\tau,{\rm out}}$.}
\label{tab:utau}
\begin{tabular}{lrllrrlll}
\hline\noalign{\smallskip}
$\eta$         &$\Rey$~~&$u_{\tau,{\rm in}}/\uin$&$u_{\tau,{\rm out}}/\uin$&$\Rey_{\tau,{\rm in}}$&$\Rey_{\tau,{\rm out}}$&$\Delta y_{\rm in}^+$&$r^+\Delta \theta$&$\Delta z_{\rm max}^+$\\ \noalign{\smallskip}\hline\noalign{\smallskip}
\multirow{2}*{0.5}& 500&0.1313&0.0656& ~32.8&16.4&0.04&1.61&~1.64\\
                  &4000&0.0823&0.0412& 164.6&82.3&0.21&4.04&~8.23\\ \noalign{\smallskip}\hline\noalign{\smallskip}
\multirow{2}*{0.4}& 500&0.1401&0.0560& ~35.0&14.0&0.05&1.15&~1.75\\
                  &4000&0.0874&0.0350& 174.9&70.0&0.11&2.86&~4.37\\ \noalign{\smallskip}\hline\noalign{\smallskip}
\multirow{2}*{0.3}& 500&0.1516&0.0454& ~37.9&11.4&0.05&0.80&~1.90\\
                  &4000&0.0941&0.0280& 188.2&55.9&0.12&1.98&~4.70\\ \noalign{\smallskip}\hline\noalign{\smallskip}
\multirow{7}*{0.2}& 500&0.1781&0.0355& ~44.6&~8.9&0.06&0.55&~2.23\\
                  &1000&0.1424&0.0284& ~71.2&14.2&0.09&0.87&~3.56\\
                  &1250&0.1336&0.0267& ~83.5&16.7&0.11&1.02&~4.18\\
                  &1375&0.1301&0.0260& ~89.4&17.8&0.11&1.10&~4.47\\
                  &1500&0.1274&0.0254& ~95.6&19.1&0.12&1.17&~4.78\\
                  &2000&0.1189&0.0238& 118.9&23.8&0.15&1.46&~5.95\\
                  &4000&0.1053&0.0211& 210.6&42.1&0.13&1.29&10.53\\
\noalign{\smallskip}\hline
\end{tabular}
\end{table}

\section{Dependences on radius ratio and Reynolds number}
\label{sec:eta}

\begin{figure*}[t]
\begin{center}
 \includegraphics[height=23mm]{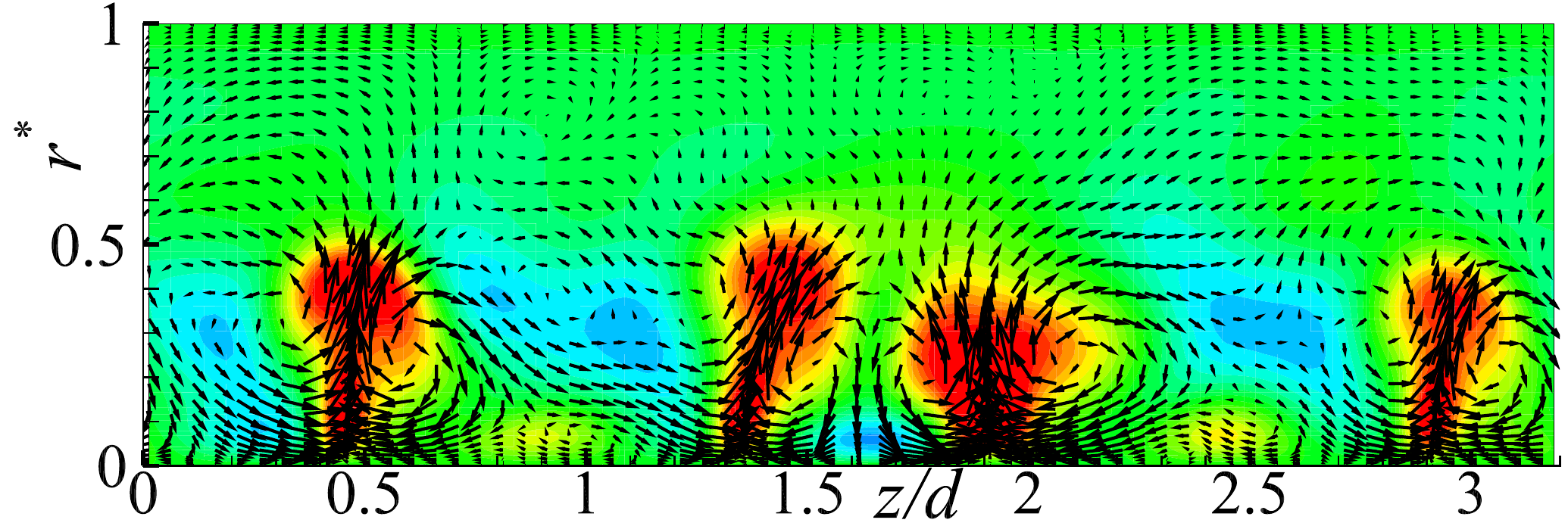} \hspace{-6mm}
 \includegraphics[height=23mm]{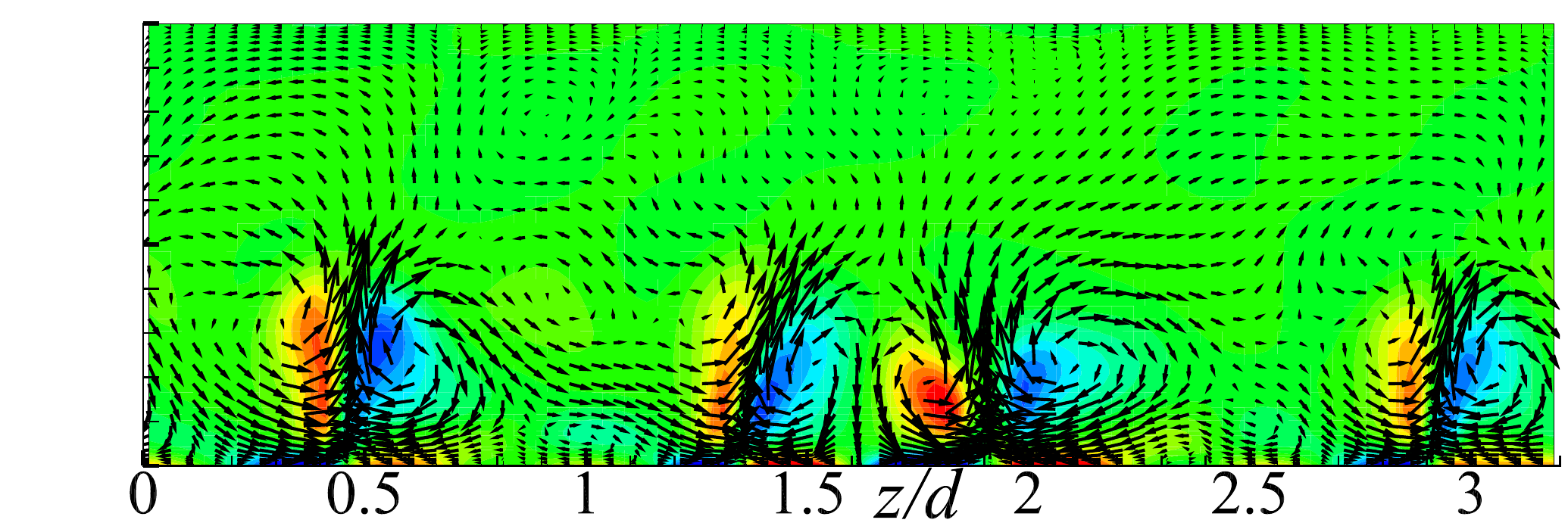}\\
 (a) $\eta=0.2$ and $\Rey=500$\\
 \vspace{1em} 
 \includegraphics[height=23mm]{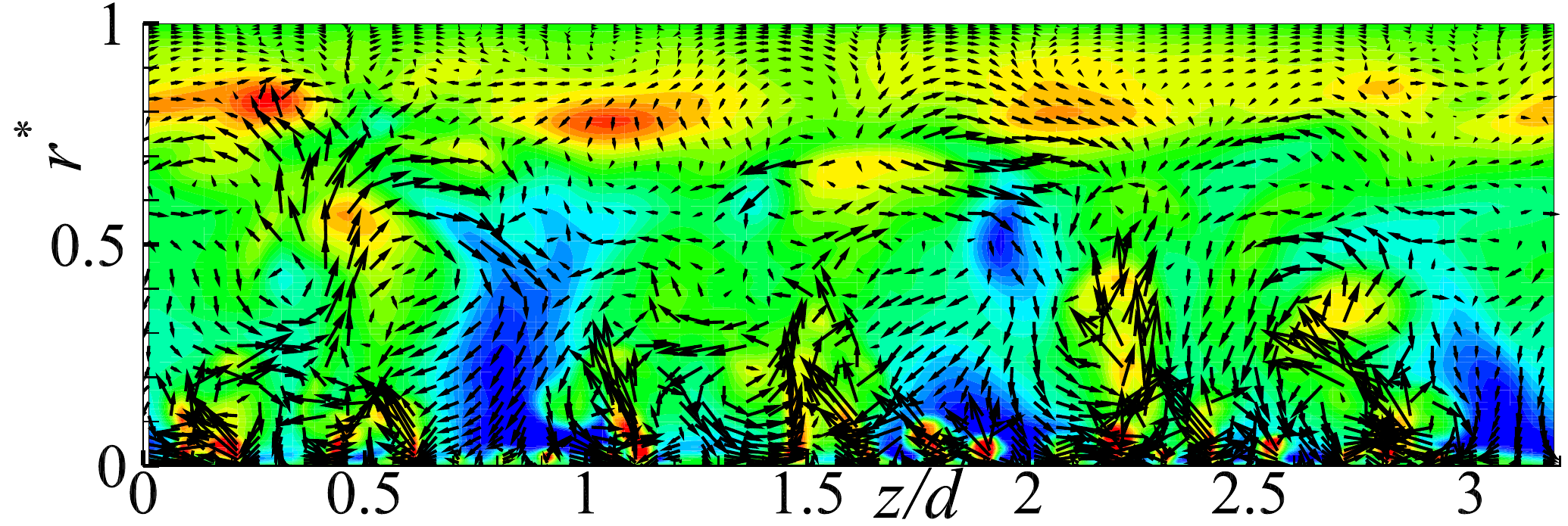} \hspace{-6mm}
 \includegraphics[height=23mm]{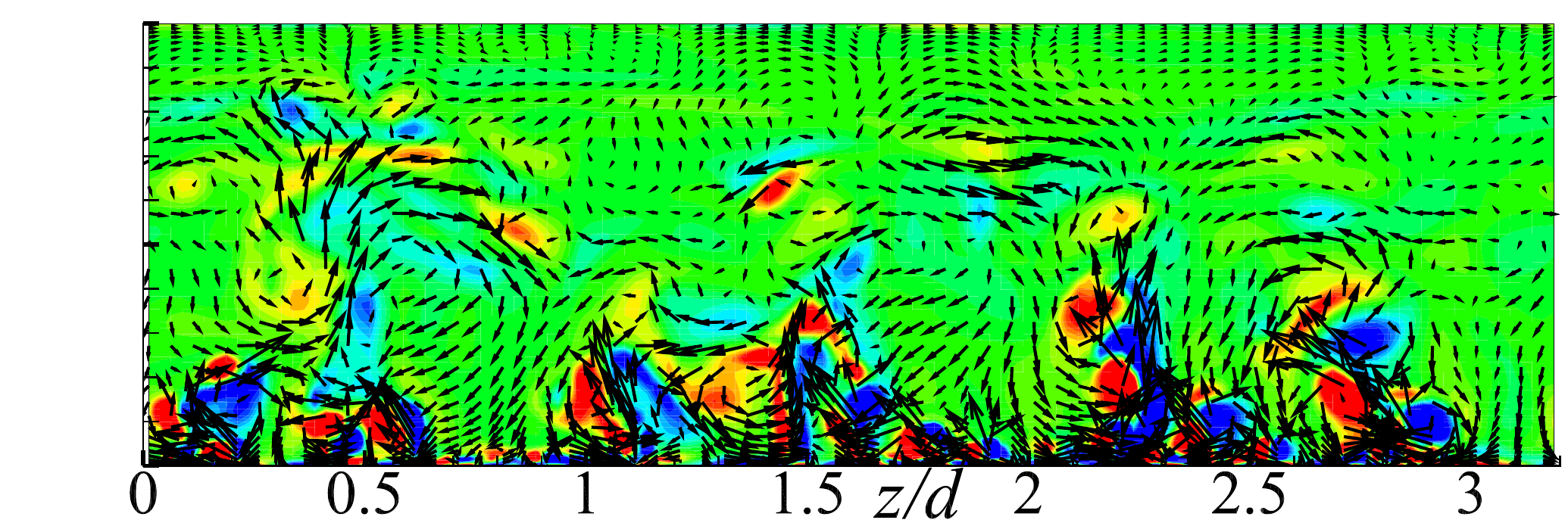}\\
 (b) $\eta=0.2$ and $\Rey=4000$\\
 \vspace{1em} 
 \includegraphics[height=23mm]{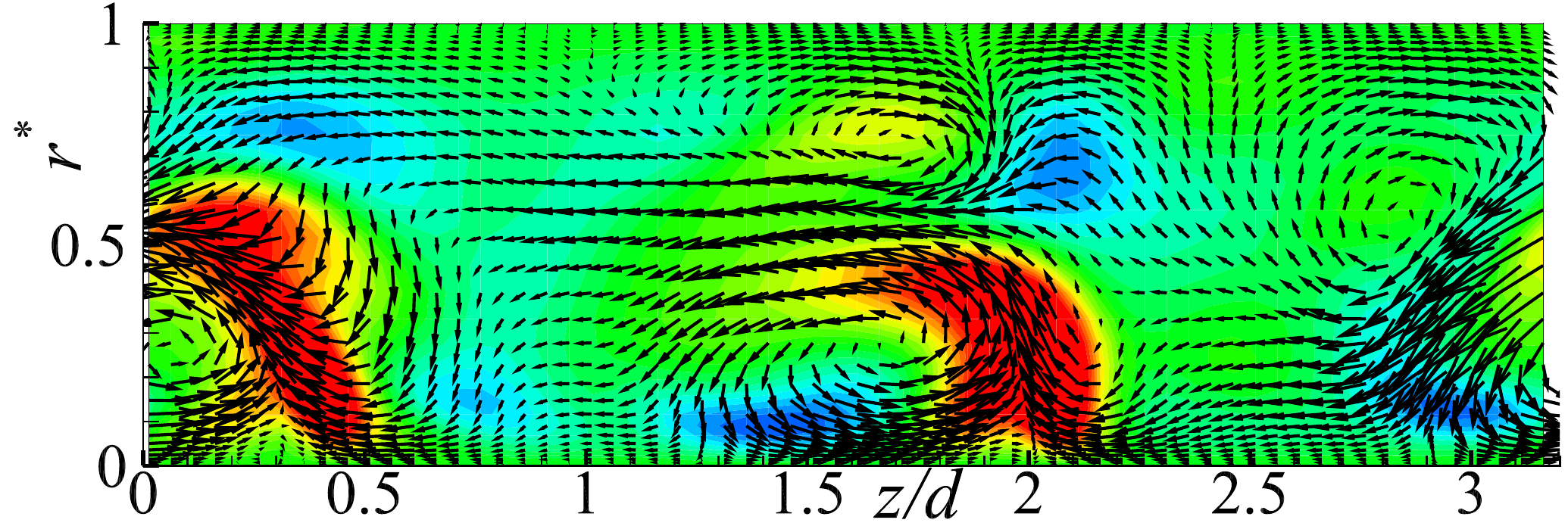} \hspace{-6mm}
 \includegraphics[height=23mm]{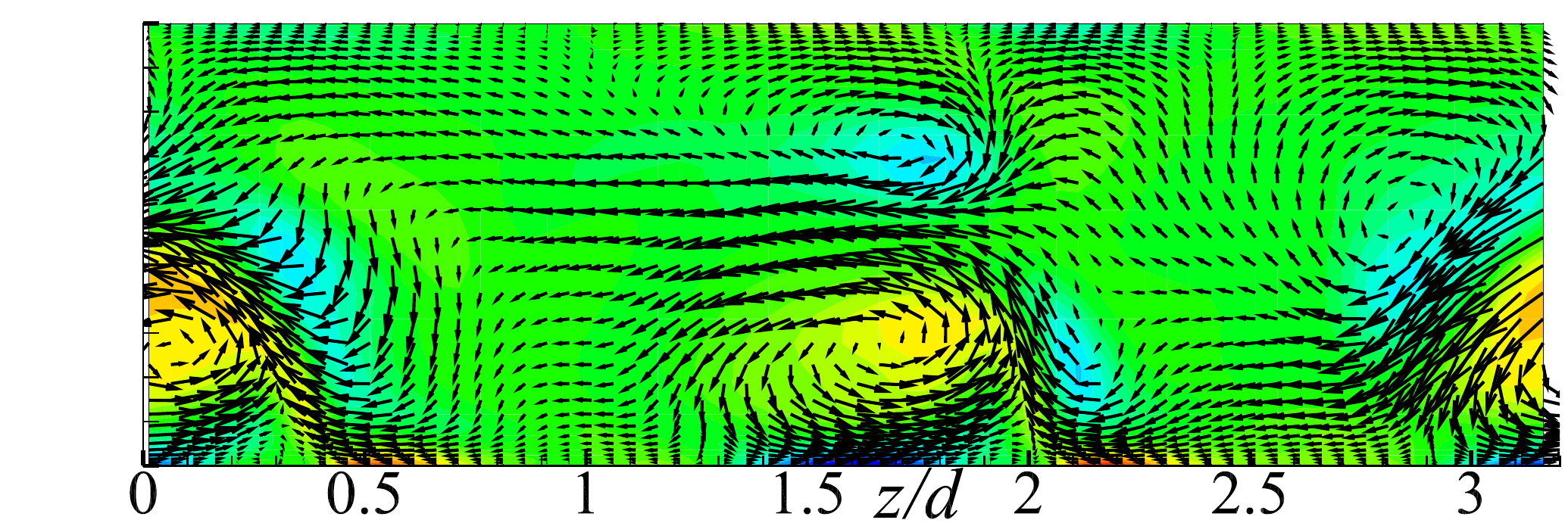}\\
 (c) $\eta=0.5$ and $\Rey=500$\\
 \vspace{1em} 
 \includegraphics[height=23mm]{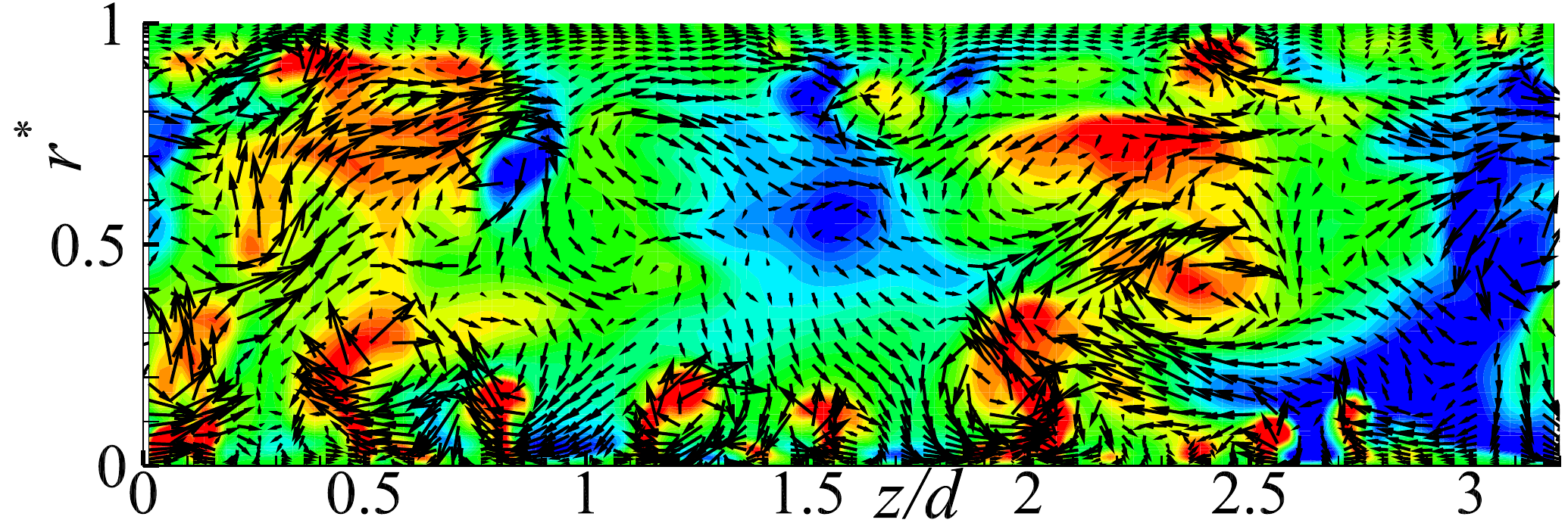} \hspace{-6mm}
 \includegraphics[height=23mm]{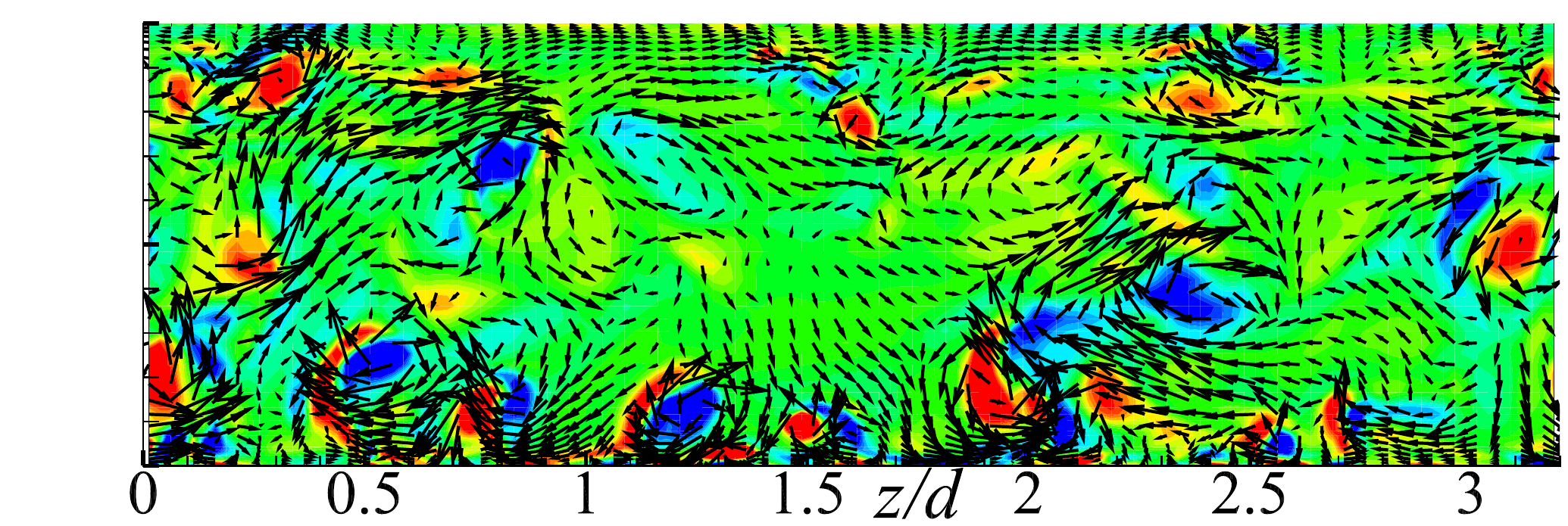}\\
 (d) $\eta=0.5$ and $\Rey=4000$\\
 \caption{Cross-sectional contour in arbitrary $z$-$r$ plane for different Reynolds numbers and radius ratios. Left column: azimuthal velocity fluctuation in the range from $u_z'/\uin = -0.2$ (blue) to 0.2 (red); right column: azimuthal vorticity in the range from $\omega_z d/\uin=-4$ (blue: clockwise rotation in figure) to 4 (red: counter-clockwise rotation in figure). The vectors show the in-plane velocity field at the same instance under each condition.}
 \label{fig:uxy}
\end{center}
\end{figure*}

In this section, we describe our investigation of the flow structure dependence on the radius ratio $\eta$ at different Reynolds numbers and report a characteristic flow structure, which can be observed only at the lowest $\eta=0.2$. 

\Fref{fig:uxy} compares the instantaneous flow fields in a cross-sectional ($z$-$r$) plane between $\eta=0.2$ and 0.5 for two Reynolds numbers. 
Note that the vertical coordinate represents the distance from the inner cylinder wall scaled by the cylinder gap (cf., \eref{eq:r*}).
The contour colours indicate the fluctuating streamwise (azimuthal) velocity $u_\theta^\prime$ or the streamwise vorticity $\omega_\theta = \partial_z u_r'-\partial_r u_z'$, while the black arrows represent the in-plane velocity vector pattern. 
For each flow, the two panels visualise the $u_\theta^\prime$ and $\omega_\theta$ at the same time instance and in the same plane. 
It can be clearly seen that the counter-rotating vortex pairs occur frequently near the inner cylinder and are accompanied by strong ejecting motions from the bottom surface. For instance, in the right panel of \fref{fig:uxy}(a), four pairs of positive and negative patches emerge in $r^* < 0.5$, while a high-speed lump moves upward at the interface of each vortex pair (see the corresponding left panel of the figure).
In both $\eta$ cases, at the lower Reynolds number of $\Rey=500$, such streamwise vortices arise owing to the centrifugal instabilities clearly observed on the inner cylinder side. However, a smaller vortex, such as a turbulent fluctuation, seems to be absent. 
On the outer cylinder side, weak streamwise vortices counter-rotating against those on the inner cylinder side can also be seen in the case of $\eta=0.5$ (see \fref{fig:uxy}(c)), while the $\eta=0.2$ case exhibits a calm field without noticeable streamwise vortices. 
This difference in the flow structure on the outer side between two $\eta$ cases can be understood by considering the flow instability on each side. 
It is obvious that on the inner cylinder side the flow is linearly unstable owing to centrifugal instabilities, whereas, in the vicinity of the outer cylinder wall, the flow (assuming there is no influence from the inner cylinder side) is linearly stable. 
In the $\eta=0.2$ case, the near outer cylinder region is not affected by the flow structure on the inner cylinder side because the cylinder gap is relatively wide in comparison to the cylinder radii. 
In the case of $\eta=0.5$, however, the outer cylinder side is affected by the inner cylinder side as a result of the relatively narrow cylinder gap. 

At the higher Reynolds number of $\Rey=4000$, small-scale streamwise vortices can be seen near the inner cylinder wall for both radius ratios. 
Large-scale structures are also observed owing to centrifugal instabilities, as shown in \fref{fig:uxy}(b) and (d). 
At $\eta=0.5$, the large-scale structures almost fill the entire gap between the cylinders, and the high/low speed regions associated with these structures reach from the inner to the outer cylinder. 
The contour of $\omega_\theta$ reveals particularly small-scale near-wall vortices only. However, a roughly vortical motion as large as half the gap width may be detected. 
Such a large-scale vortical motion that dominates even at high Reynolds numbers is known as the Taylor-vortex. 
Then, almost the entire field with $\eta = 0.5$ is in a state of the Taylor-vortex flow. 
However, as is the case of $\eta=0.2$, the spatial radial extent of the Taylor-votex reaches $r^* \approx 0.6$ at most, in contrast to the larger $\eta$ case. 
\Fref{fig:uxy}(b) manifests the outer cylinder side with a rather homogeneous structure in the spanwise ($z$) direction, and also the absence of the streamwise vortex even at the high value of $\Rey=4000$. 
It should be noted that another field visualized at different $\theta$ and/or time instance may demonstrate a similar spanwise-homogeneously expanded structure, but with a negative fluctuation $u'_\theta$.

\begin{figure}[t]
\begin{center}
\includegraphics[width=0.666\hsize]{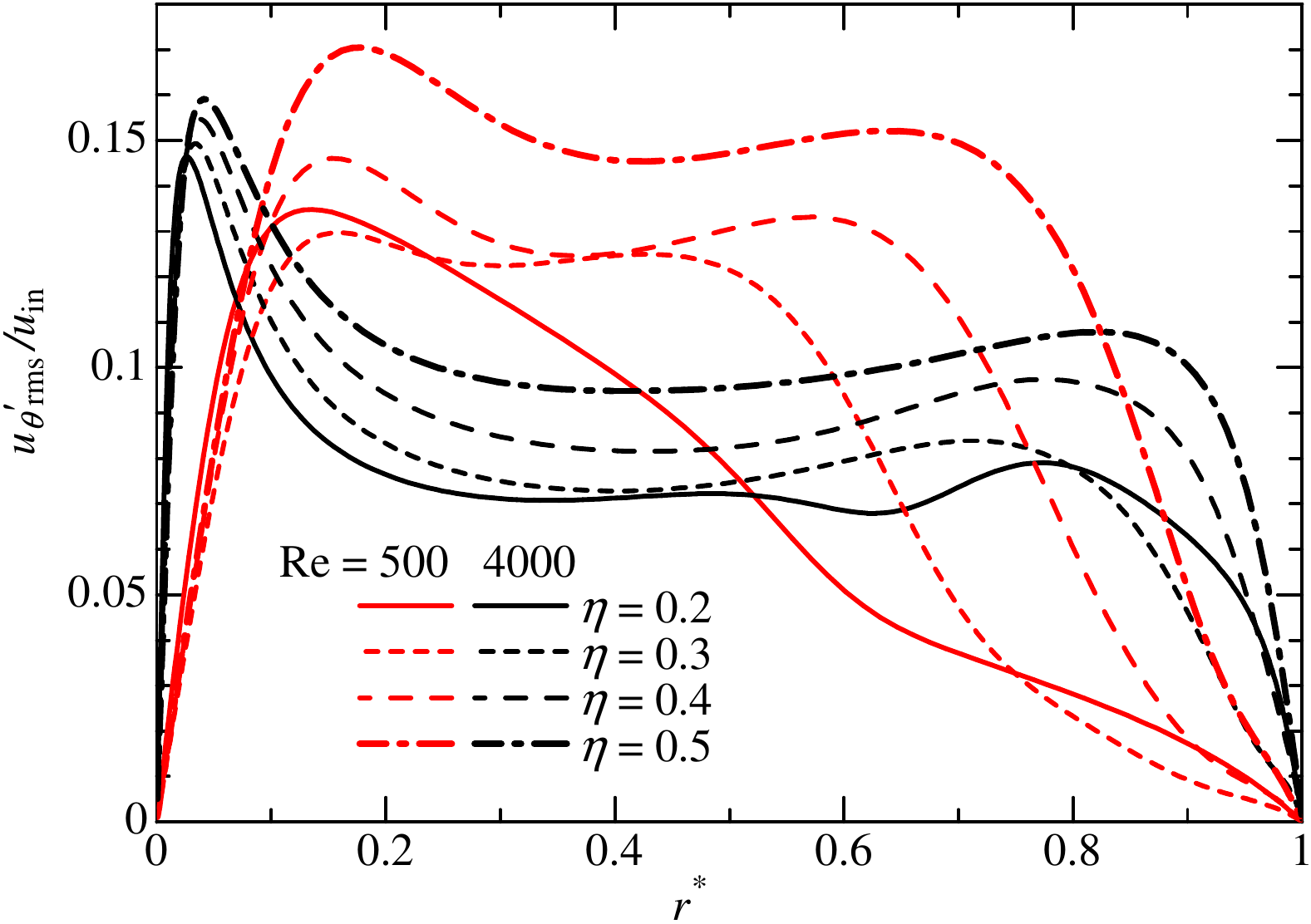}
\caption{Root-mean-square profiles of azimuthal velocity fluctuation $u'_\theta$ scaled by wall speed $u_\mathrm{in}$ at (red) $\Rey=500$ and (black) $\Rey=4000$ for different radius ratios.}
\label{fig:etawrms}
\end{center}
\end{figure}

The above-mentioned difference in the flow structure between the small and the moderate radius ratio cases can be confirmed in the profiles of several turbulent statistics. 
\Fref{fig:etawrms} shows the wall-normal (radial) profiles of the root-mean-square of the streamwise (azimuthal) velocity fluctuation for a different $\eta$ at the two Reynolds numbers. 
At the lower value of $\Rey=500$, the $u_\theta^\prime$ profile for $\eta=0.2$ indicates a qualitatively different tendency in comparison to those in the larger $\eta$ cases. 
For $0.3 \leq \eta$, the profile has two peaks on each of the inner and outer sides. 
It can also be seen that the magnitude decreases and the outer peak location moves towards the inner side (in terms of $r^\ast$) as the $\eta$ decreases, i.e., as the cylinder gap becomes wider.  
The inner and outer peaks of the velocity fluctuation correspond to the near-wall vortices, such as those observed in \fref{fig:uxy}(c) and (d). 
At $\eta=0.2$, the outer peak disappears and $u'_{\theta \rm{,rms}}$ decreases monotonically with the increasing radial position $r^\ast$ on the outer side. 
This corresponds to the velocity field shown in \fref{fig:uxy}(a), where a vortical structure is not observed on the outer side. 
It is also noteworthy that the inner peak magnitude at $\eta=0.2$ is larger than that at $\eta=0.3$, although for $0.3 \leq \eta \leq 0.5$, the inner peak magnitude decreases as the $\eta$ decreases. 

Additionally, at the higher value of $\Rey=4000$, the $u_\theta^\prime$ profile in the lowest $\eta$ case of $\eta=0.2$ indicates a qualitatively different tendency from the larger $\eta$ cases, as indicated by the black curves in \fref{fig:etawrms}. 
For $\eta \geq 0.3$, the profiles have a sharp peak near the inner cylinder wall, and in the central part and outer side of the cylinder gap the velocity fluctuation is rather constant, but increases slightly in the vicinity of $r^\ast=0.7$--0.8. 
The inner peaks correspond to the small-scale structure near the inner cylinder wall observed in \fref{fig:uxy}(d), and the rather-constant $u_\theta^\prime$ distribution in the central and outer region of the cylinder gap is caused by the large-scale structures including the Taylor vortices that fill up almost the entire gap between the cylinders. 
However, the profile at $\eta=0.2$ has an extra broad peak in the vicinity of $r^\ast=0.8$, which corresponds to the spanwise-homogeneously expanded structure (called the `modal structure' later) near the outer cylinder wall observed in \fref{fig:uxy}(b), in addition to an apparent peak at $r^\ast=0.5$. 
The local-minimum position at $r^\ast=0.63$ may be considered as a demarcation between the inner side Taylor-vortex flow and the outer side modal structure.

\begin{figure}[t]
\begin{center}
\includegraphics[width=0.666\hsize]{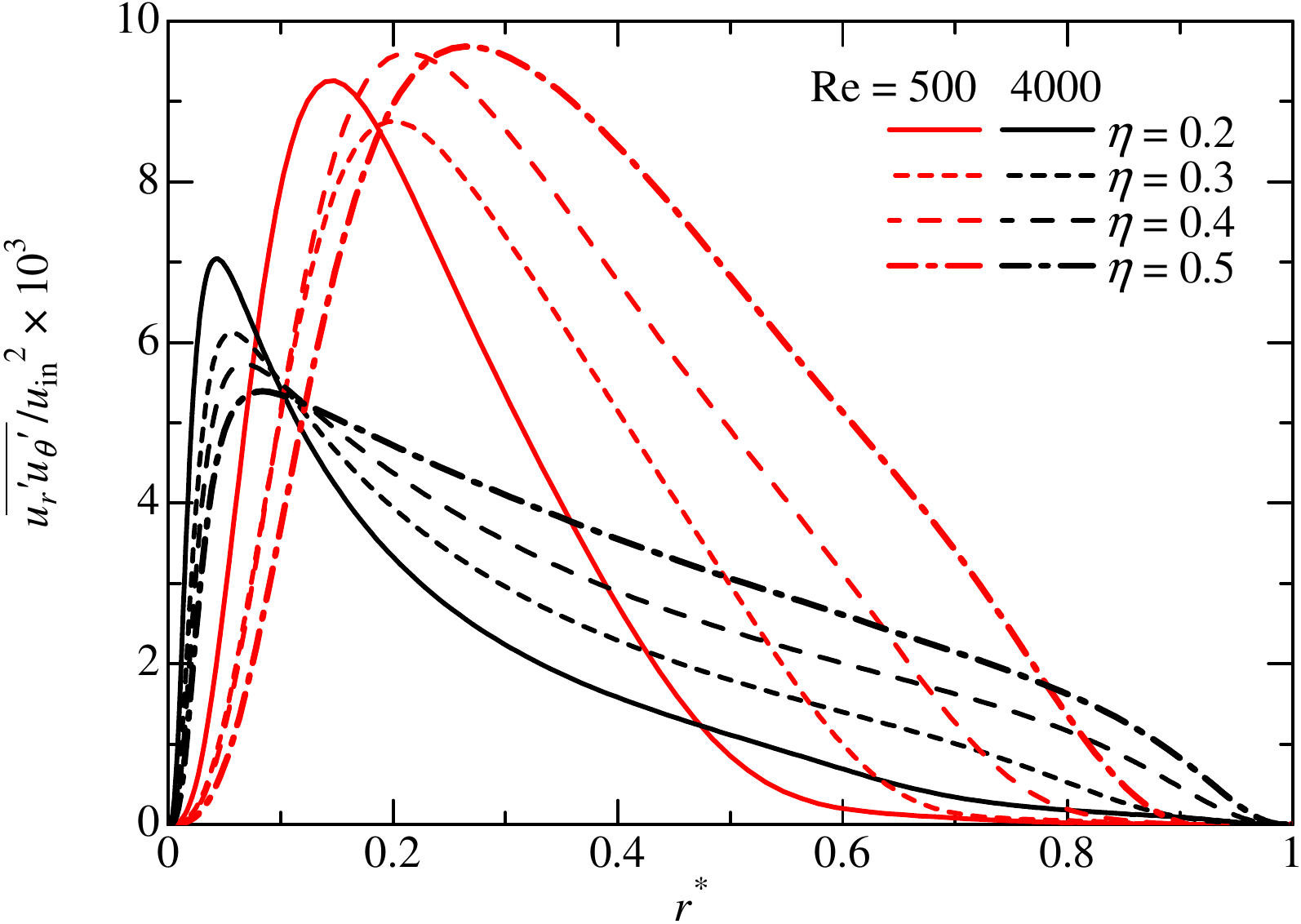}
\caption{Profiles of Reynolds shear stress $\overline{u'_r u'_\theta}$ scaled by $u_\mathrm{in}^2$ for different radius ratios. The colours and lines indicate the same as in \fref{fig:etawrms}.}
\label{fig:etarss}
\end{center}
\end{figure}

\Fref{fig:etarss} presents the wall-normal profiles of the Reynolds shear stress $\overline{u_r^\prime u_\theta^\prime}$. 
Since $\partial_r \overline{u_\theta} < 0$, the sign of $\overline{u_r^\prime u_\theta^\prime}$ should be positive throughout the entire gap. 
Even in the approximately laminar flow at $\Rey=500$, the magnitude of $\overline{u_r^\prime u_\theta^\prime}$ is large in the central region of the cylinder gap, which implies that a significant momentum transfer is induced by the Taylor-vortex. 
This is the reason that the peak is located on the inner side in every case. 
The more prominent peak for $\Rey=4000$ is also caused by the centrifugal instability on the inner cylinder, which produces small-scale vortical structures and ejecting motions, as visualized in \fref{fig:uxy}. 
\Fref{fig:etarss} further reveals that the peak moves more or less toward the inner cylinder as the $\eta$ decreases. 
At $\eta=0.2$, the Reynolds shear stress is almost zero on the outer half of the cylinder gap. 
Additionally, in the turbulent case of $\Rey=4000$, the Reynolds shear stress changes linearly at the central part of the cylinder gap at $\eta=0.5$. As the $\eta$ decreases, the magnitude of the peak located near the inner cylinder wall increases and the profile in the central part of the cylinder gap becomes increasingly concave.  
It is interesting to note that the outer side modal structure does not contribute to the Reynolds shear stress, but exhibits velocity fluctuations of significant magnitude (cf., \fref{fig:etawrms}).

\begin{figure*}[t]
\begin{center}
\begin{tabular}{ll}
\begin{minipage}{0.35\hsize}
\begin{center}
 \includegraphics[height=70mm]{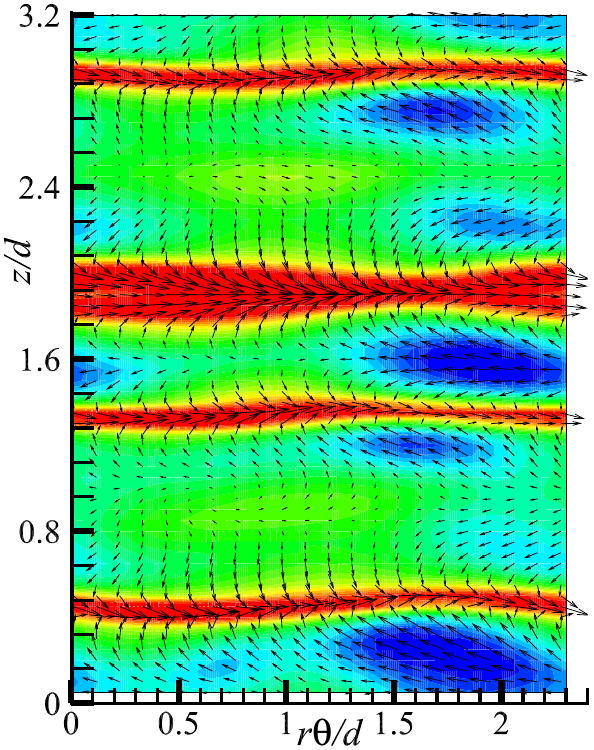}\\
 (a) $\eta=0.2$, $\Rey=500$
\end{center}
\end{minipage}
\begin{minipage}{0.645\hsize}
\begin{center}
 \includegraphics[height=70mm]{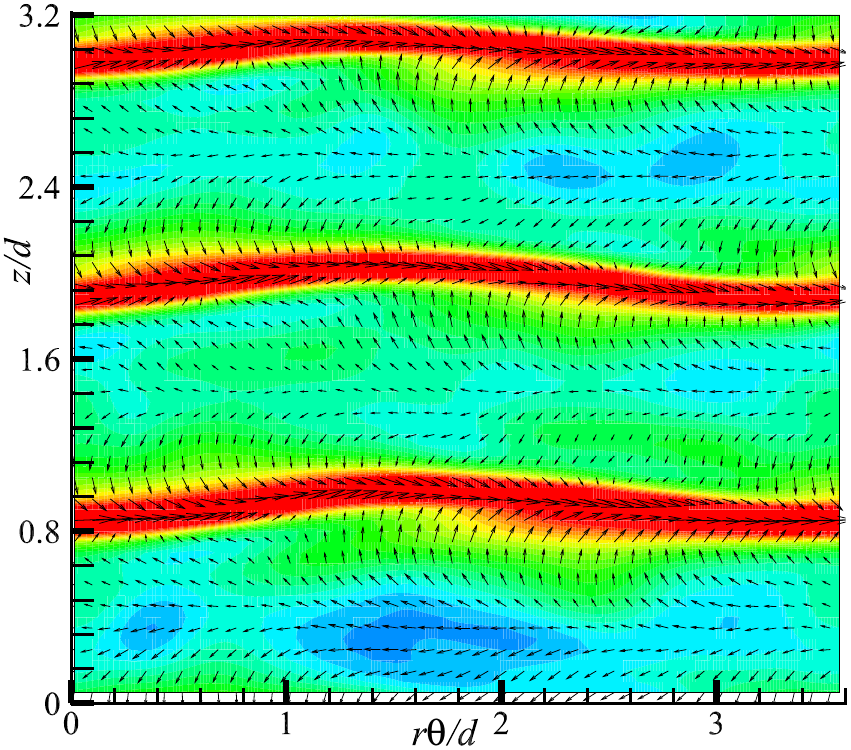}\\
 (b) $\eta=0.3$, $\Rey=500$
\end{center}
\end{minipage}
\end{tabular}
\vspace{2em} \\
\begin{tabular}{cc}
\begin{minipage}{0.35\hsize}
\begin{center}
 \hspace{-5mm}
 \includegraphics[height=70mm]{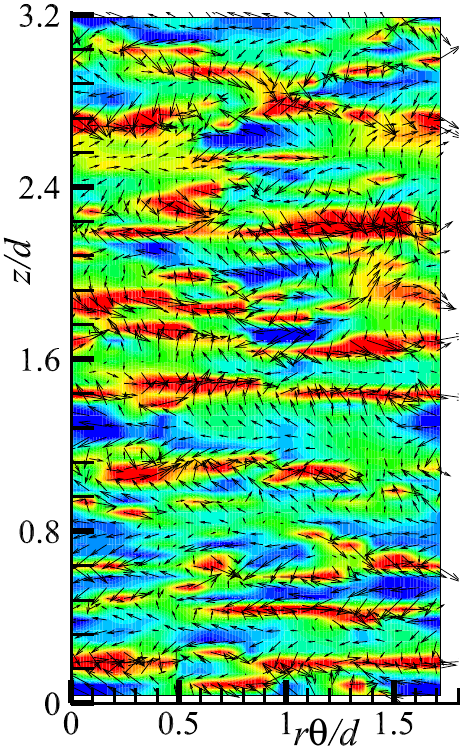}\\
 (c) $\eta=0.2$, $\Rey=4000$
\end{center}
\end{minipage}
\begin{minipage}{0.645\hsize}
\begin{center}
 \includegraphics[height=70mm]{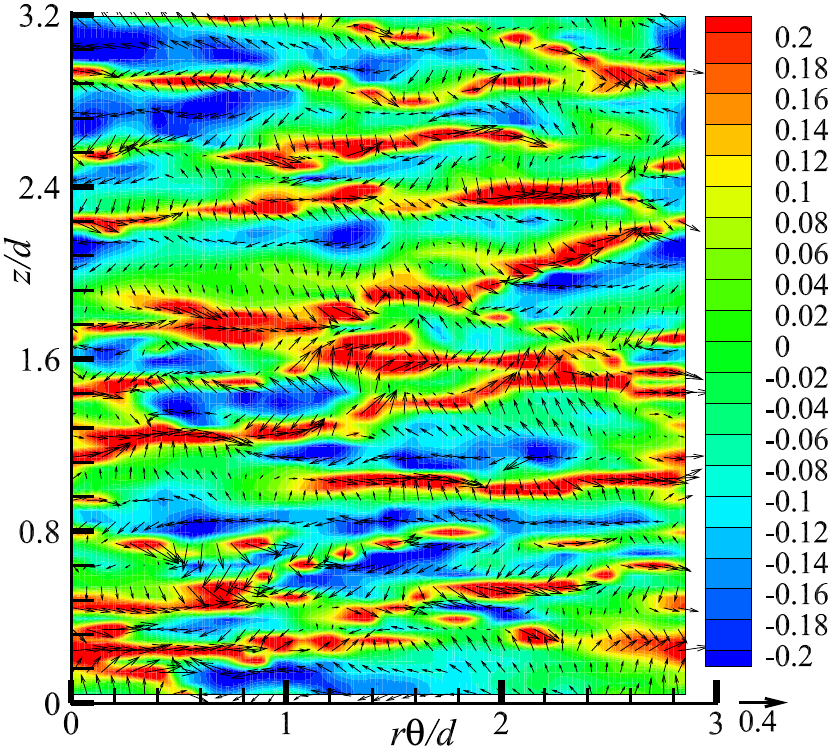}\\
 (d) $\eta=0.3$, $\Rey=4000$
\end{center}
\end{minipage}
\end{tabular}
\end{center}
\caption{Instantaneous velocity fluctuation fields in the $\theta$-$z$ plane at $y_{\rm in}^+=10.1$--10.6 in the vicinity of inner-cylinder surface. The contour shows the azimuthal velocity component $u'_\theta/\uin$, and the vectors represent in-plane velocities ($u'_\theta, u'_z$) normalized by $\uin$; the colour range and reference vector are shown in (d). The entire field of $0\leq z \leq L_z$ and $0\leq \theta \leq 2\pi$ is visualized.}
\label{fig:uixz}
\end{figure*}

\begin{figure*}[t]
\begin{tabular}{ll}
\begin{minipage}{0.42\hsize}
\begin{center}
 \includegraphics[height=30mm]{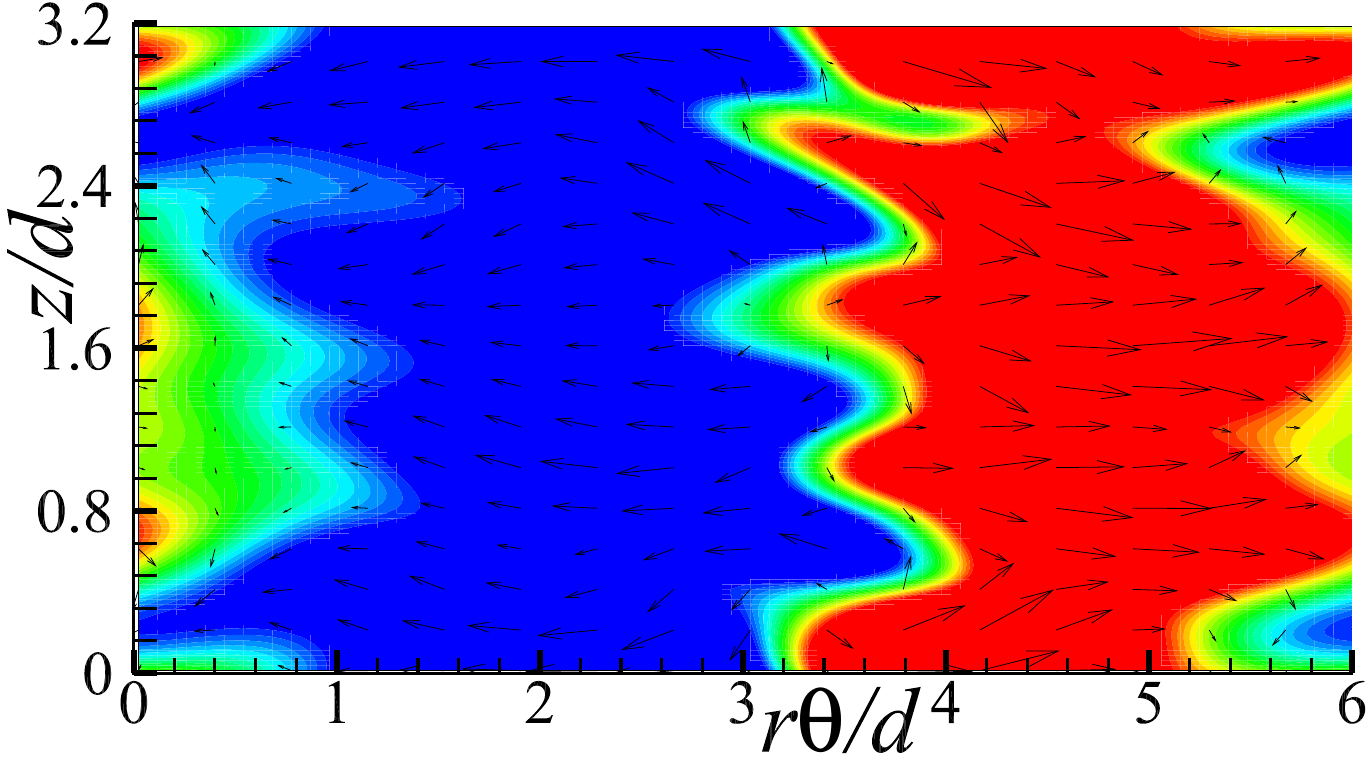}\\
 (a) $\eta=0.2$, $\Rey=500$
\end{center}
\end{minipage}
\begin{minipage}{0.58\hsize}
\begin{center}
 \includegraphics[height=30mm]{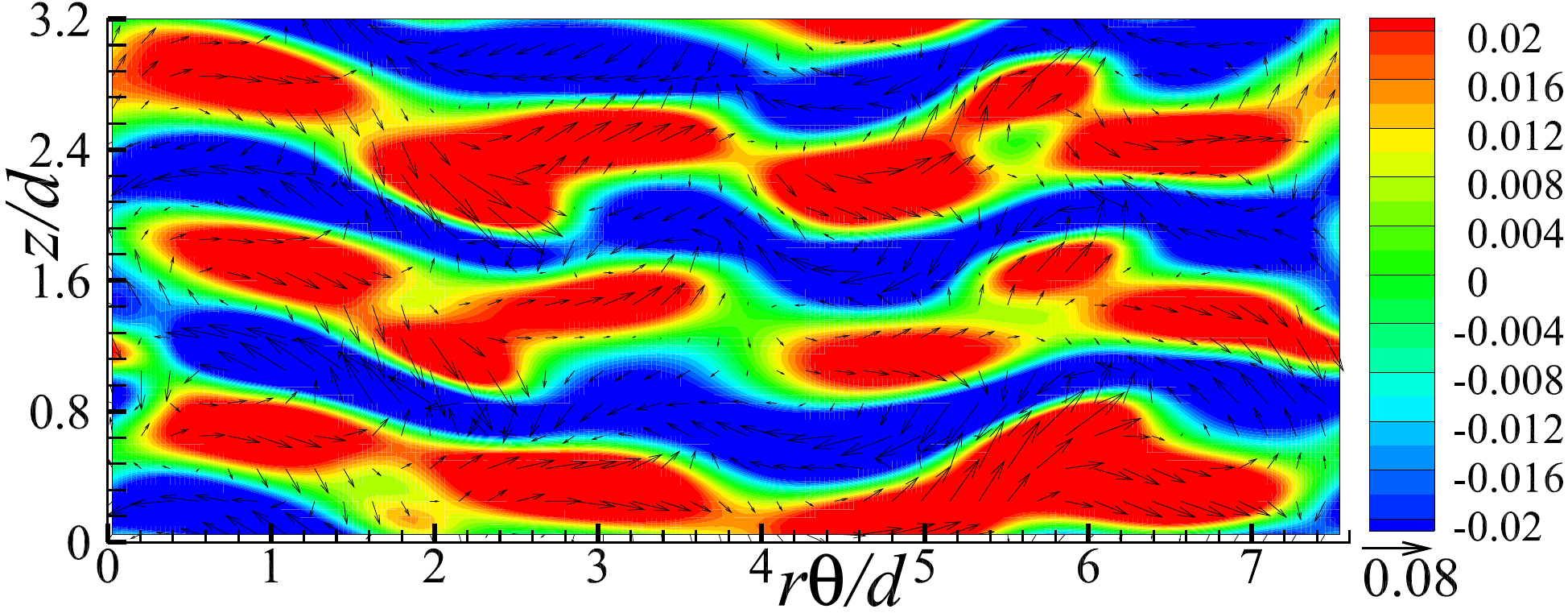}\\
 (b) $\eta=0.3$, $\Rey=500$
\end{center}
\end{minipage}
\end{tabular}
\vspace{0.5em} \\
\begin{tabular}{cc}
\begin{minipage}{0.42\hsize}
\begin{center}
 \includegraphics[height=30mm]{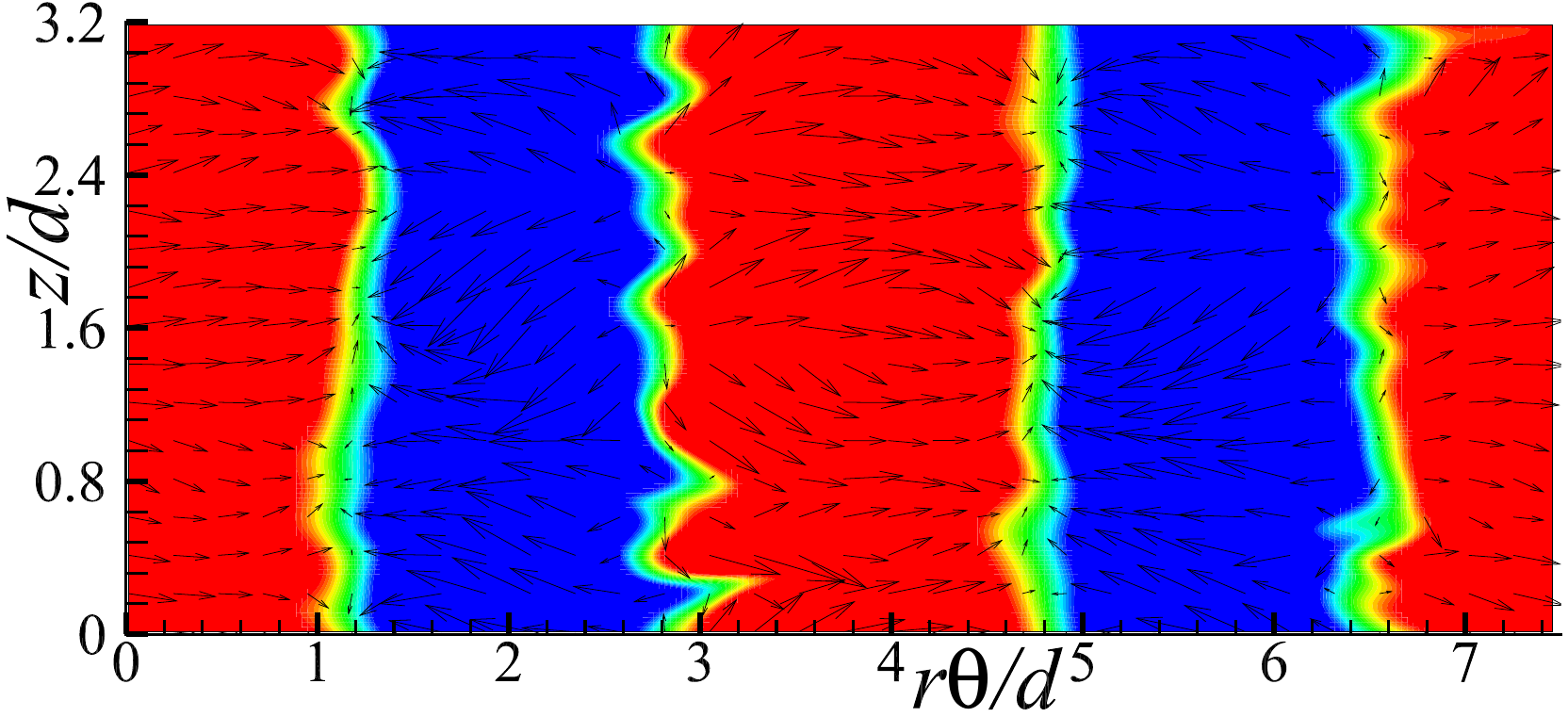}\\
 (c) $\eta=0.2$, $\Rey=4000$
\end{center}
\end{minipage}
\begin{minipage}{0.58\hsize}
\begin{center}
 \includegraphics[height=30mm]{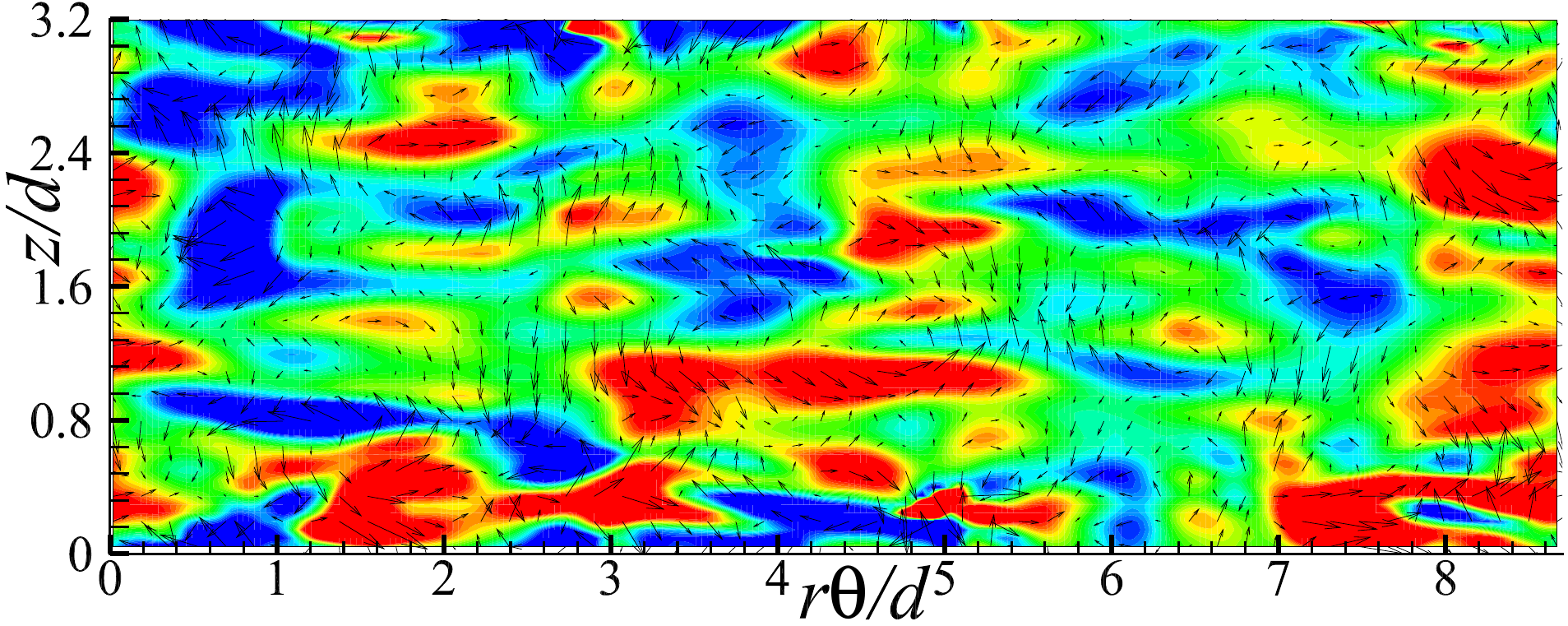}\\
 (d) $\eta=0.3$, $\Rey=4000$
\end{center}
\end{minipage}
\end{tabular}
\caption{Same as \fref{fig:uixz}, but at $y_{\rm out}^+=5.13$--5.21 in the vicinity of the outer-cylinder surface. The colour range and reference vector are shown in (b).}
\label{fig:uoxz}
\end{figure*}

As described above, there must be a critical point of the radius ratio between $\eta=0.2$ and 0.3, below which the flow structure on the outer cylinder side clearly changes from the Taylor-vortex flow. 
To perceive the structural change more closely, the instantaneous flow fields in the vicinity of the inner and outer cylinder wall are visualized in Figs.~\ref{fig:uixz} and \ref{fig:uoxz}. 
The radial positions of the visualized $\theta$-$z$ planes are unified in $y^+ = y u_\tau/ \nu \approx 10$ or 5 in the wall unit from each cylinder surface, where $y_{\rm in}$ and $y_{\rm out}$ are the wall-normal distances from the inner and outer cylinders, respectively, and either $u_{\tau, {\rm in}}$ or $u_{\tau, {\rm out}}$ is used for the corresponding side.
In the figures, the two slightly different $\eta$ cases of $0.2$ and 0.3 are compared, but the modal structure that exists only on the outer side of $\eta=0.2$ can be easily observed.
First, let us focus on the inner cylinder side.
As shown in \fref{fig:uixz}, the flow near the inner-cylinder wall does not show any significant difference in the flow structure between the two $\eta$ cases for $\Rey=500$ and 4000. In the approximately laminar case of $\Rey=500$, the coherent structure elongated in the streamwise (azimuthal) direction is observed for both $\eta=0.2$ and 0.5, as shown in \fref{fig:uixz}(a, b). In the turbulent case, \fref{fig:uixz}(c, d) shows high/low-speed streaks corresponding to the near-wall turbulent structures. 
The regularly arranged streaks that are common for \fref{fig:uixz}(a, b) provide the evidence for the existence of the Taylor-vortices, and the slightly-meandering streaks relate to the wavy Taylor-vortex flow.

On the outer cylinder side, a clear difference can be seen in the structure between $\eta=0.2$ and 0.3. 
As shown in \fref{fig:uoxz}(b) with regard to the case of $\Rey=500$ and $\eta=0.3$, streamwise-elongated structures similar to those on the inner side are observed on the outer-cylinder side. 
In comparison with the inner side structure shown in \fref{fig:uixz}(b), it can be seen that the spanwise width (in $z$) and the wavelength of the structure’s streamwise meandering, which is observed on the outer side shown in \ref{fig:uoxz}(b), are consistent with those on the inner cylinder side. 
Moreover, some additional instabilities are also observed in the spatial velocity variation in the streamwise direction on the outer side. 
Thereby, it may be conjectured that such a flow structure observed near the outer cylinder wall is a footprint of the wavy Taylor-vortex on the inner side. 
However, such streamwise elongated streaks are not observed for $\eta=0.2$. 
Instead, \Fref{fig:uoxz}(a) exhibits a band of negative $u'_\theta$ in an azimuthal extent of $r\theta/d = 0$--3, and another band of positive $u'_\theta$ in 3--6. 
It should be noted that this pattern propagates at approximately mean velocity on the outer-half of the gap without any change in shape, except for the wavy interface of the bands. 
The band interface is wavy and its wavelength apparently coincides with the spanwise spacing of the Taylor vortices at the inner cylinder. 
A careful comparison with \fref{fig:uixz}(a) may reveal the interaction of the structures between the inner and outer sides. However, this tendency becomes fuzzy at a higher $\Rey$.
As shown in \fref{fig:uoxz}(d), the footprint of the inner side structure is not as clearly observed on the outer cylinder side, even for $\eta=0.3$. 
In the smaller $\eta=0.2$, the axially-homogeneous modal velocity variations remain at $\Rey=4000$, in \fref{fig:uoxz}(c). 
The difference between the low- and high-$\Rey$ cases observed here is in the azimuthal mode (i.e., wavenumber) of the band patterns; that is, the mode is one for $\Rey=500$ and shifts to two for $\Rey=4000$, as illustrated by \fref{fig:uoxz}(a, c).

\section{Modal structure}
\label{sec:mode}

\begin{figure}[t]
\begin{center}
\includegraphics[width=0.666\hsize]{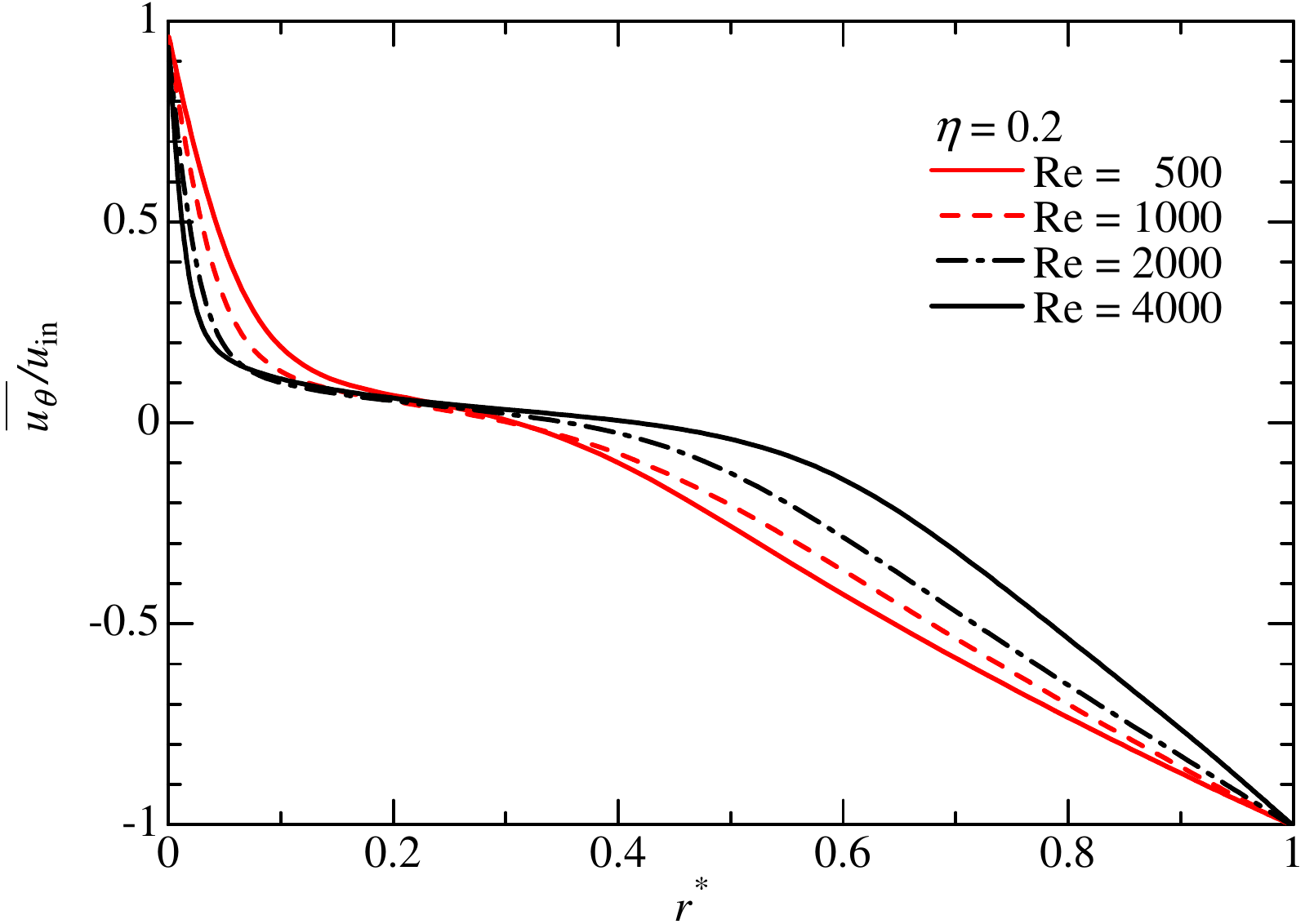}
\caption{Mean azimuthal velocity $\overline{u_\theta}$ for different Reynolds numbers.}
\label{fig:Rewmean}
\end{center}
\end{figure}

As described in the previous section, at the small radius ratio of $\eta=0.2$, a spanwise-homogeneous modal structure exists stably on the outer cylinder side and the azimuthal mode number of such a pattern depends on the Reynolds number. 
We have observed that the modal structure is not stationary but rather moves in the azimuthal direction with the outer cylinder wall. 
These characteristics resemble those seen in the Tollmien-Schlichting (TS) wave. 
It would be worthwhile to attempt to consider the analogy of the plane Poiseuille flow with spanwise system rotation, which is known as the rotating channel flow and consists of both a linearly unstable side and a stable side owing to the Coriolis force effect, similar to the current flow system. 
On the stable side of the rotating channel flow, the TS wave is triggered by the disturbance from the unstable side and forms stably, as has been demonstrated by Brethouwer et al.~\cite{RefG}. 
Accordingly, we conjecture that, in the current flow system, the velocity disturbance by the coherent structures on the inner cylinder side could give rise to the TS wave on the stable outer cylinder side and result in the currently observed modal structure. 

Now, we will focus on the smallest radius ratio case of $\eta=0.2$ and investigate the Reynolds number dependency of the flow structure on the outer cylinder side in detail. 
\Fref{fig:Rewmean} presents the mean azimuthal velocity profiles for $\eta=0.2$ at different Reynolds numbers from $\Rey=500$ to 4000. 
As shown in the figure, the mean velocity decreases monotonically as $r^\prime$ increases from $\uin$, and each profile has a plateau of $\overline{u_\theta} \approx 0$ on the inner cylinder side. 
The plateau expands as much as $r^* = 0.1$--0.5 for $\Rey=4000$.
As the Reynolds number increases, the mean velocity gradient also increases in the near-wall regions on both sides

\begin{figure}[t]
\begin{center}
\includegraphics[width=0.666\hsize]{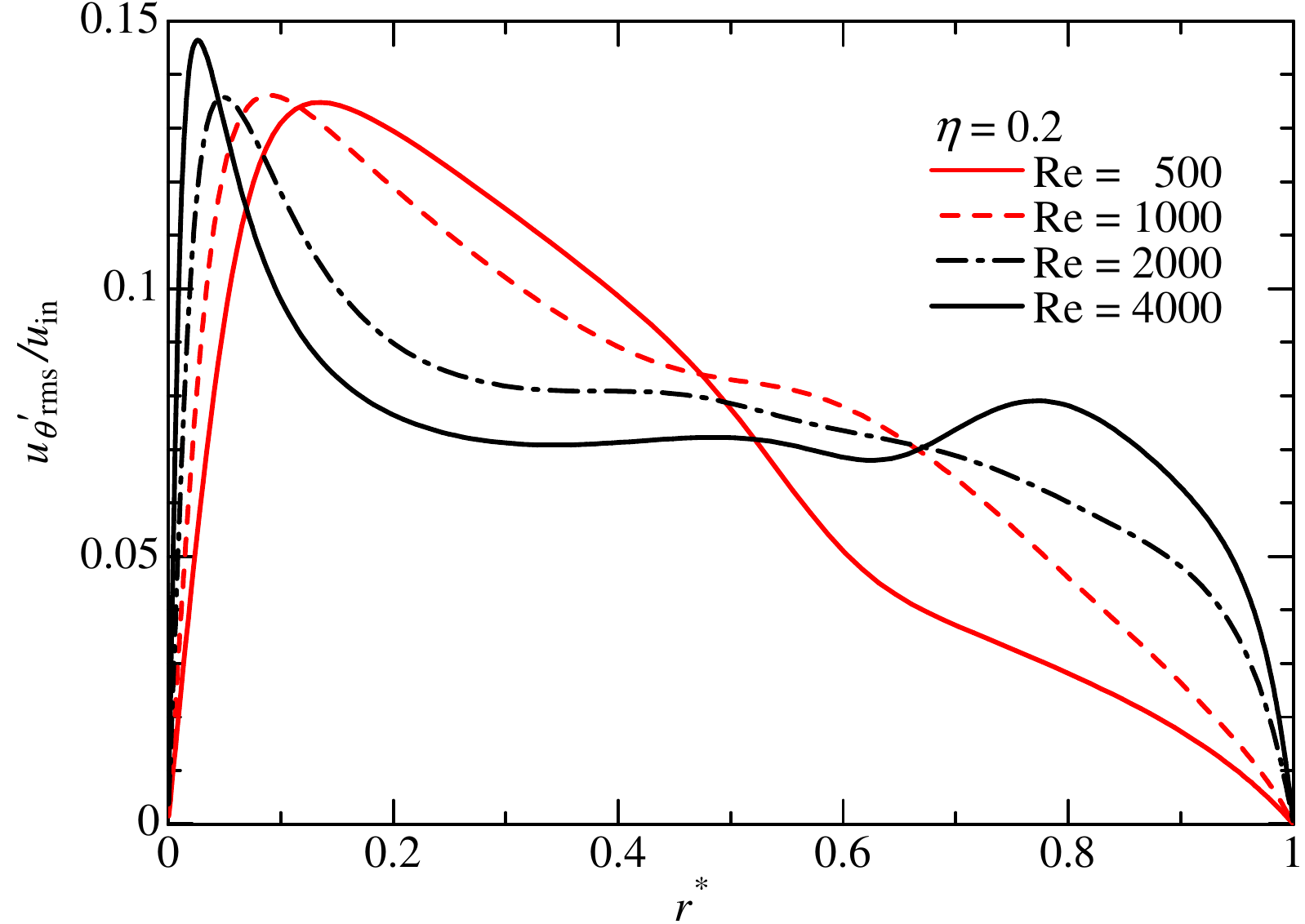}
\caption{Root-mean-square of $u'_\theta$ for different Reynolds numbers at $\eta=0.2$.}
\label{fig:Rewrms}
\end{center}
\end{figure}

\begin{figure}[t]
\begin{center}
\includegraphics[width=0.666\hsize]{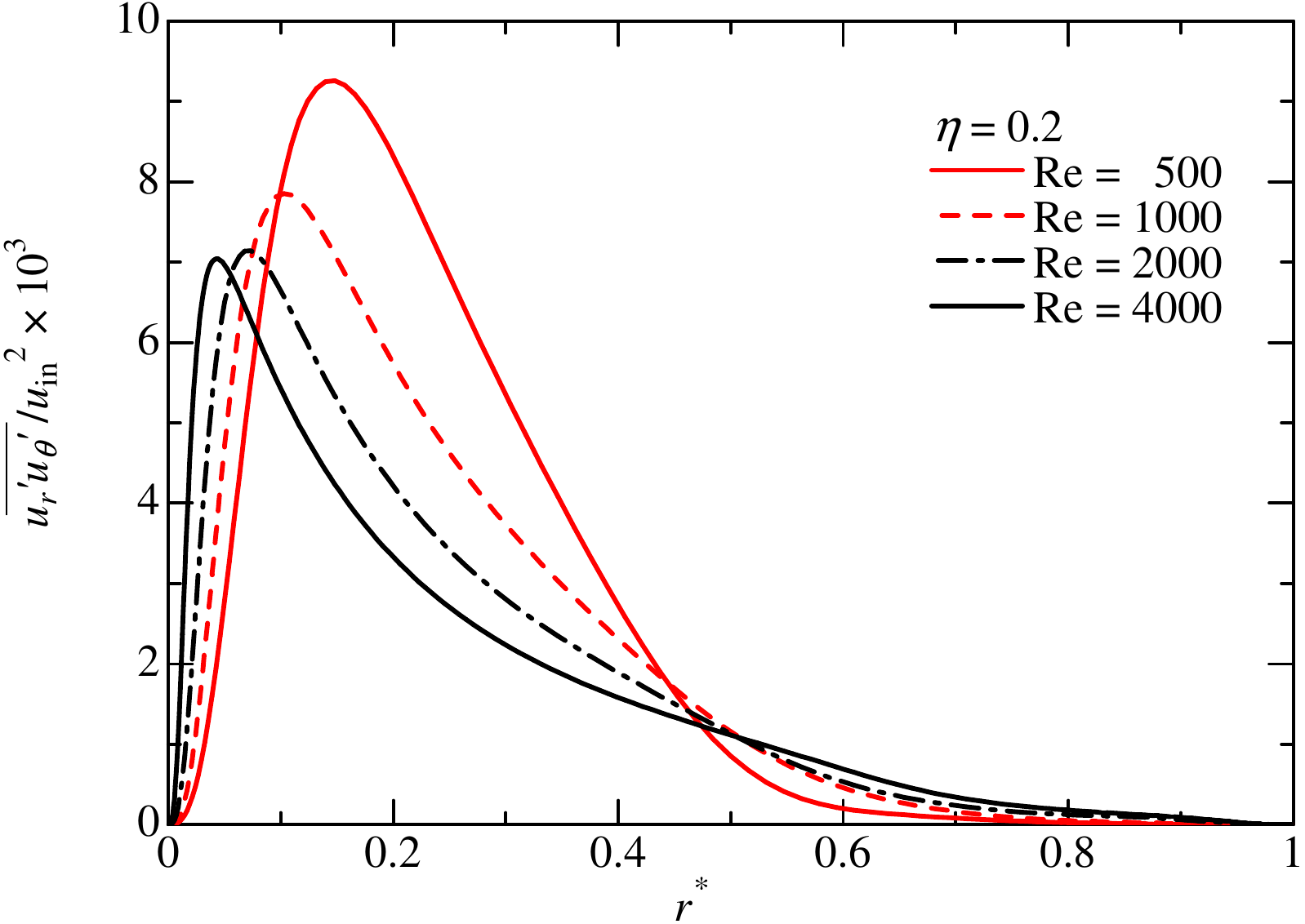}
\caption{Reynolds shear stress $\overline{u'_r u'_\theta}$ for different Reynolds numbers at $\eta=0.2$.}
\label{fig:Rerss}
\end{center}
\end{figure}

In \fref{fig:Rewrms}, the profiles of the streamwise velocity fluctuation intensity $u'_{\theta, {\rm rms}}$ for $\eta=0.2$ are also compared for different Reynolds numbers. 
As shown here, the peak near the inner cylinder increases in magnitude and moves closer towards the inner cylinder wall, as the Reynolds number increases. 
This inner peak at a high $\Rey$ corresponds to the near-wall turbulence structures. 
On the outer side, $u'_{\theta, {\rm rms}}$ similarly increases as $\Rey$ increases, while its magnitude is moderately relative to the inner one. 
It is noteworthy that the outer peak at $r^\ast=0.8$, which appears at $\Rey=4000$ (as already observed in \fref{fig:etawrms}), does not exist at  smaller Reynolds numbers. 
This may indicate a structural change between $\Rey=2000$ and 4000.
However, such a sudden change at a high Reynolds number cannot be detected in the profile of the Reynolds shear stress. 
As reported in \fref{fig:Rerss}, the $\overline{u'_r u'_\theta}$ profile has a peak on the inner side, whereas, on the outer cylinder side, the profile approaches zero asymptotically. 
The Reynolds shear stress is produced only on the inner side, where the Taylor-vortex dominates the flow. 
As mentioned earlier, the modal structure might not contribute to the radial momentum transport. 

\begin{figure}[t]
\begin{center}
\includegraphics[width=130mm]{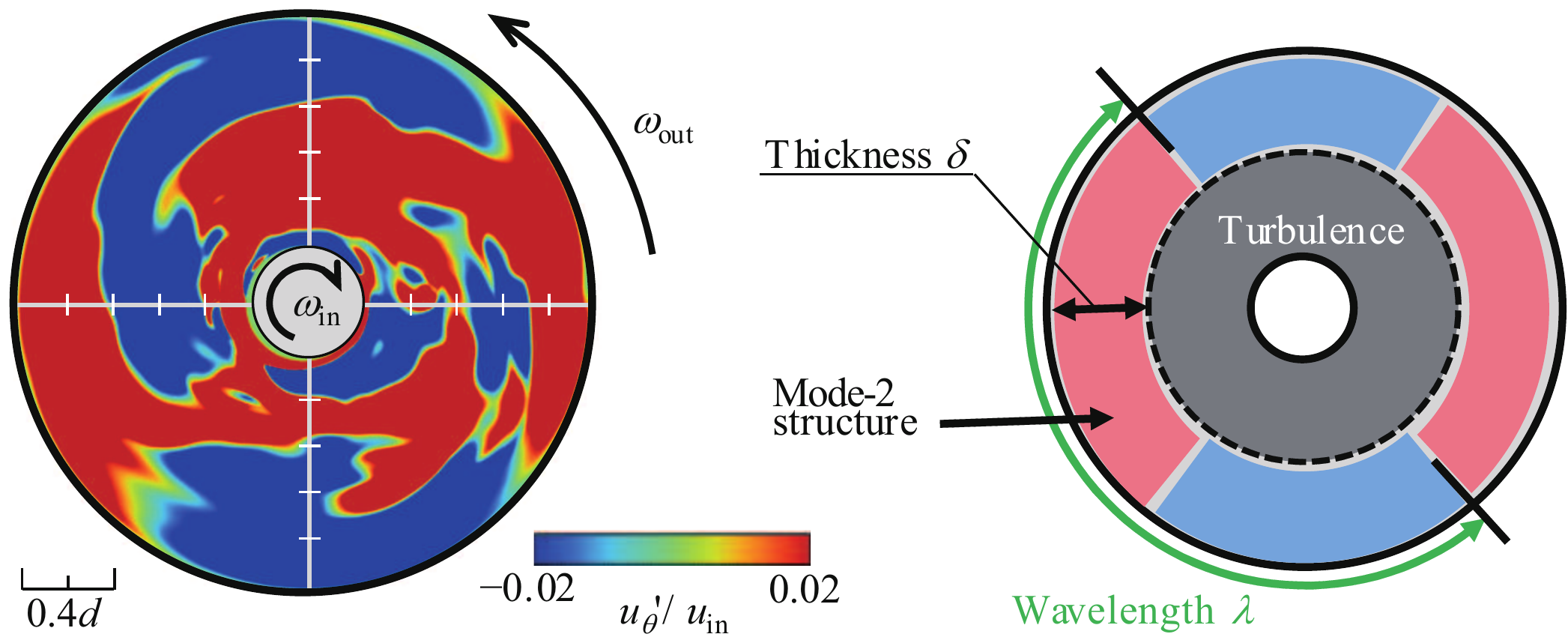}
\caption{Investigation of outer cylindrical structure. The left side shows the explanation of the global flow structure in the $r$-$\theta$ plane for $\Rey = 4000$. The counter shows the azimuthal velocity fluctuation. The right side shows the explanatory drawing of the thickness of structure $\delta$ and the wavelength in the azimuthal direction $\lambda$.}
\label{fig:TSwave}
\end{center}
\end{figure}

As has been shown above, two essentially different flow structures exist on the inner and outer cylinder sides for $\eta=0.2$. 
\Fref{fig:TSwave} presents the top view of the instantaneous velocity field between the concentric cylinders, where it can be observed that the flow structures near the inner cylinder and on the outer side are qualitatively different. 
Additionally, a roughly-interpreted schematic view is also shown. 
The flow structure on the inner side is rather close to the wall turbulence including the Taylor-vortex, while that on the outer side is a modal structure with a certain thickness $\delta$ and a well-defined wavelength $\lambda$. 
Below, we attempt to estimate the $\delta$ and $\lambda$ of the modal structure.

\begin{figure}[t]
\begin{center}
\begin{minipage}{0.495\hsize}
\begin{center}
 \includegraphics[height=44mm]{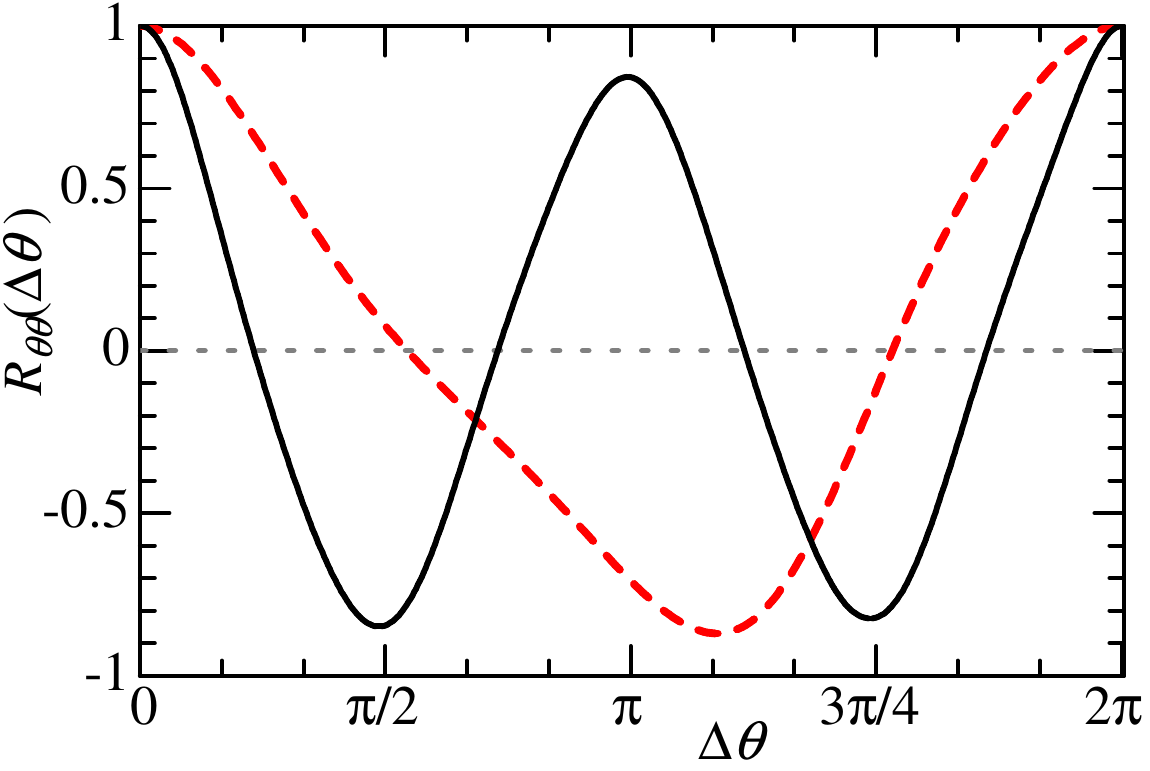}\\
 (a) Azimuthal correlation
\end{center}
\end{minipage}
\begin{minipage}{0.495\hsize}
\begin{center}
 \includegraphics[height=44mm]{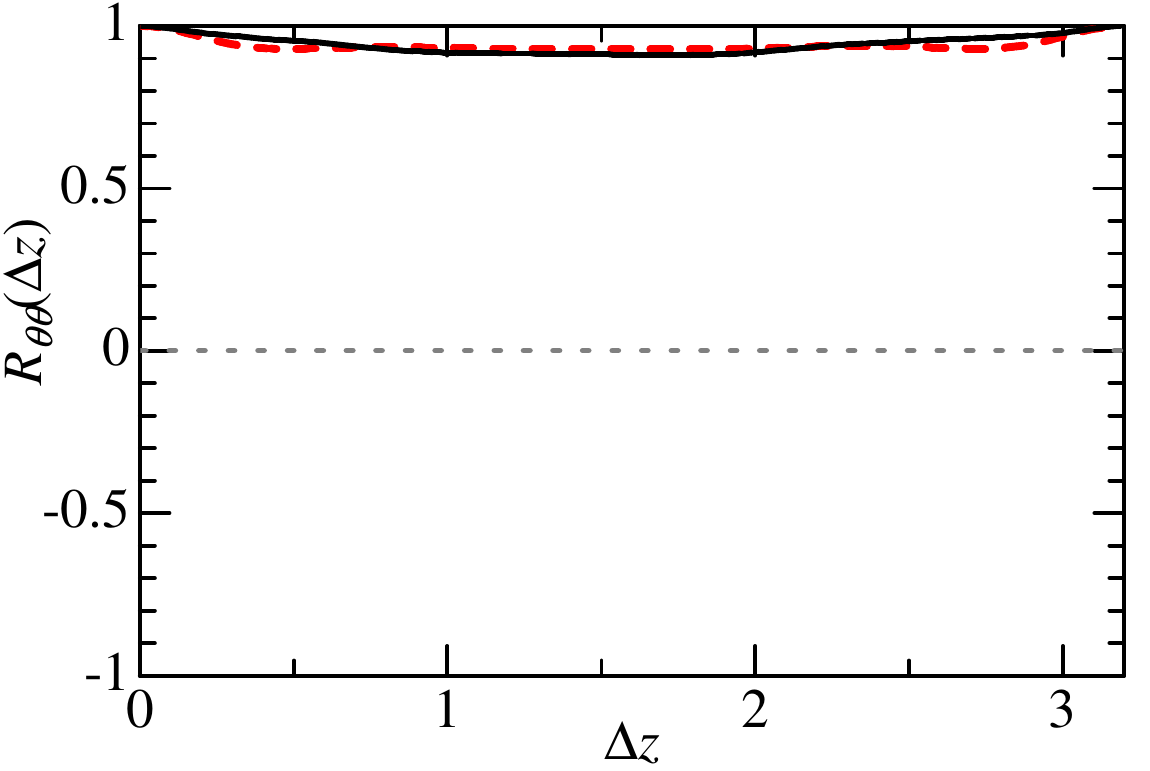}\\
 (b) Axial correlation
\end{center}
\end{minipage}
\end{center}
\caption{Two-point autocorrelation functions near the outer cylinder of $\eta=0.2$ as a function of either (a) $\Delta \theta$ or (b) $\Delta z$. The red-dashed and black solid lines in each panel represent $\Rey = 1000$ and 4000, respectively.}
\label{fig:tpc}
\end{figure}

First, $\lambda$, which is the azimuthal wavelength of the modal structure, is estimated by the two-point autocorrelation function of the fluctuating azimuthal velocity. 
\Fref{fig:tpc} presents the streamwise (azimuthal) and spanwise (axial) two-point correlation functions. 
For instance, the former is defined as follows:
\begin{equation}
R_{\theta \theta}(\Delta \theta )=
\frac{\overline{u'_\theta (r, \theta, z) \cdot u'_\theta (r, \theta+\Delta \theta, z)}}{u'_{\theta, {\rm rms}}(r) u'_{\theta, {\rm rms}}(r)}.
\label{eq:tpc}
\end{equation}
The reference radial position $r$ is in the vicinity of the outer cylinder wall. In each figure, two Reynolds-number cases with $\eta=0.2$ are compared. 
The azimuthal correlation shown in \fref{fig:tpc}(a) displays the periodicity with an approximately sinusoidal curve, and the wavelength of the azimuthal periodic pattern is $2\pi$ and $\pi$ at $\Rey=1000$ and 4000, respectively. 
Hereafter, we term the modal structures with a period of $2\pi$ and $\pi$ as `mode-1' and `mode-2', respectively. 
\Fref{fig:tpc}(b) confirms the axial (spanwise) homogeneity of the modal structure regardless of the mode and Reynolds number. 

\begin{figure}[t]
\begin{tabular}{cc}
\begin{minipage}{0.5\hsize}
\begin{center}
 \hspace{-5mm}
 \includegraphics[width=66mm]{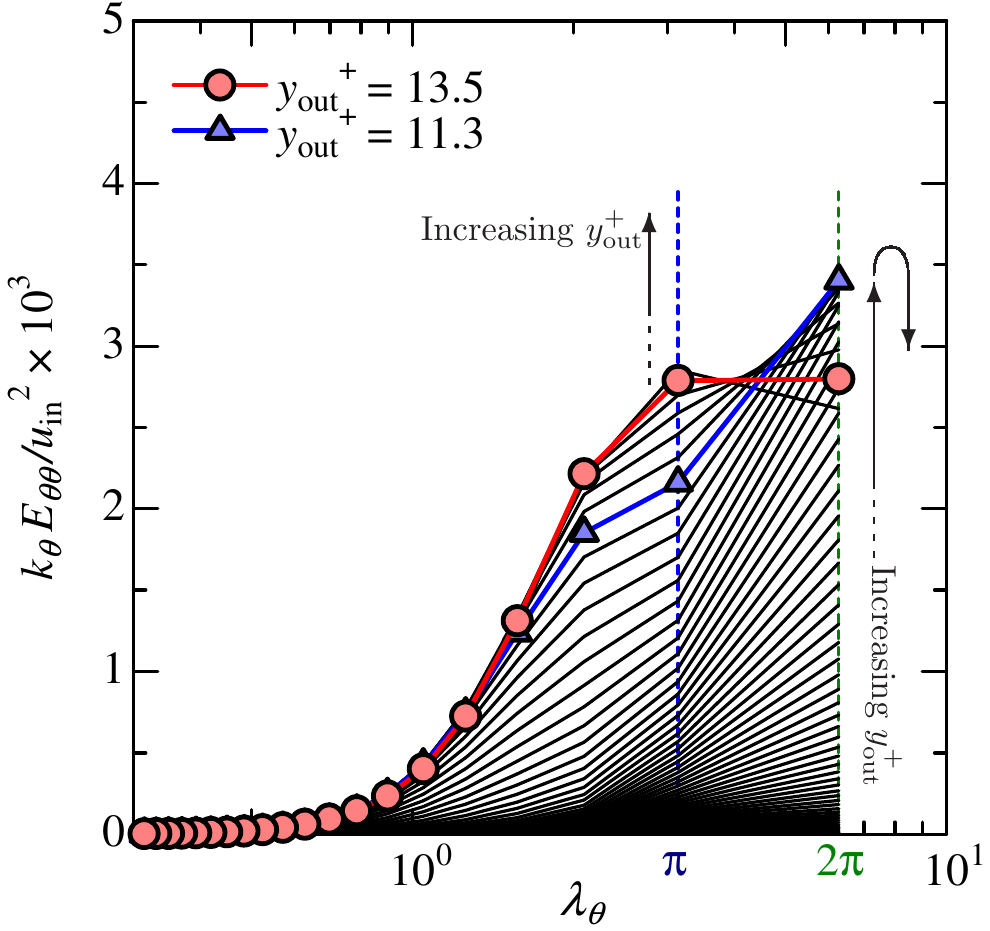}\\
 (a) $\Rey=1000$ with mode-1
\end{center}
\end{minipage}
\begin{minipage}{0.5\hsize}
\begin{center}
 \hspace{-5mm}
 \includegraphics[width=66mm]{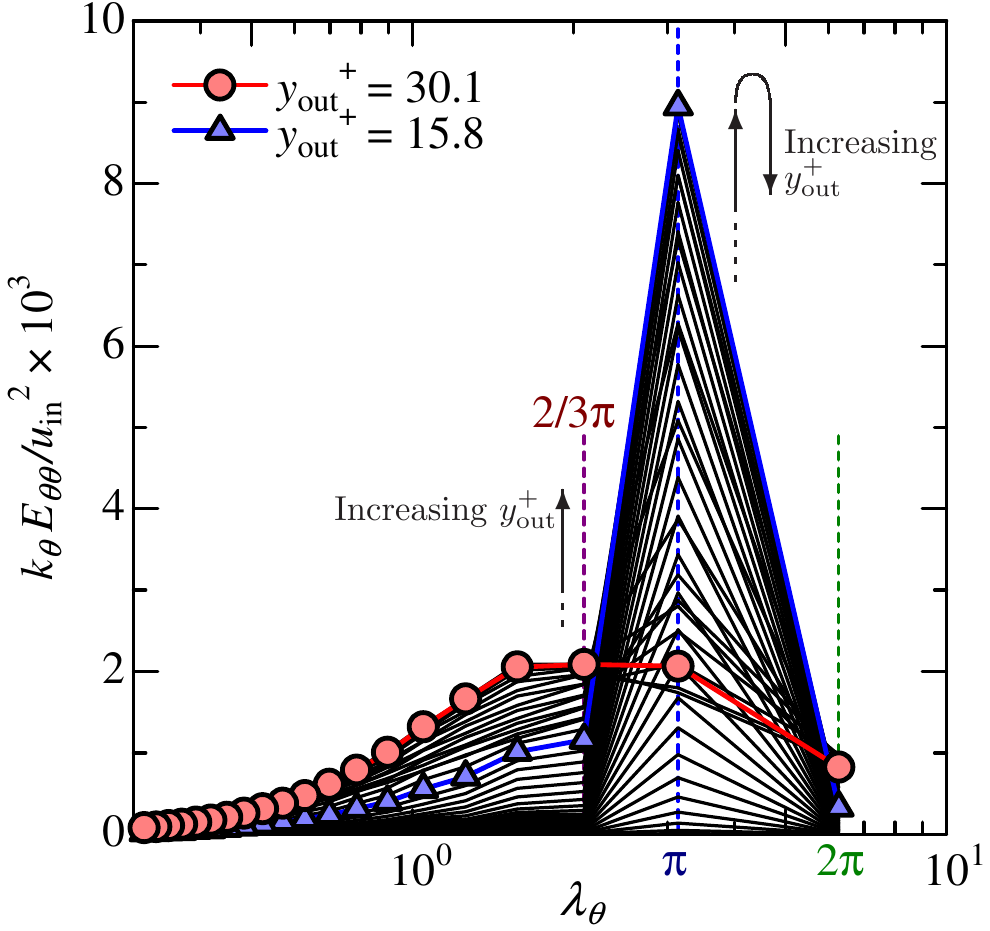}\\
 (b) $\Rey=4000$ with mode-2
\end{center}
\end{minipage}
\end{tabular}
\caption{Pre-multiplied energy spectrum $k_{\theta} E_{\theta \theta}$ of azimuthal velocity fluctuation $u'_\theta$ as a function of $\lambda_{\theta}$ at several $r_{\rm out}^+$ representing the wall-normal height from the outer cylinder in the wall-unit.}
\label{fig:delta}
\end{figure}

Secondly, we determine the radial thickness of the modal structure $\delta$, based on the azimuthal energy spectra. 
\Fref{fig:delta} shows the pre-multiplied energy spectrum $k_\theta E_{\theta \theta}$ at every grid distance from the outer cylinder wall. 
The panels (a) and (b) are $\Rey=1000$ and 4000, respectively; mode-1 and mode-2 were observed on the outer cylinder side. 
Note that $y_\mathrm{out}^+$ stands for the distance from the outer cylinder wall scaled by the viscous wall units; that is, $y_\mathrm{out}^+ \equiv (r_{\rm out} -r)u_{\tau, \mathrm{out}}/\nu$. Please refer to \tref{tab:utau} for $u_{\tau, \mathrm{out}}$.
In \fref{fig:delta}(a), the energy spectra at $y_\mathrm{out}^+=11.3$ and 13.5 are highlighted by blue and red, respectively. 
Up to $y_\mathrm{out}^+=11.3$, the energy spectra have a peak at $\lambda_\theta= 2 \pi$, which is consistent with the observation of mode-1 at $\Rey=1000$, as shown in \fref{fig:tpc}. Additionally, the peak keeps increasing as the wall-normal distance $y_\mathrm{out}^+$ increases. 
Beyond $y_\mathrm{out}^+=11.3$, the energy of $\lambda_\theta=2 \pi$ starts to decrease, while that of the smaller wavelengths keeps increasing. 
The spectrum at $y_\mathrm{out}^+=13.5$ has a broad peak consisting of two neighbouring wavelengths with comparable energy.
A further increase in $y_\mathrm{out}$ results in a peak shift towards the shorter (or high-frequency) structures inherent in the turbulence, although they are not shown in the figure.
Based on this change in the $y_\mathrm{out}^+$ dependency of the energy spectrum, we deduce that there is a structural change between $y_\mathrm{out}^+=11.3$ and 13.5. 
Then, the thickness of the mode-1 structure for $\Rey=1000$ scaled by the viscous units is estimated as $\delta^+_\mathrm{out} = \delta u_{\tau, \mathrm{out}}/\nu=13.5$. 
Additionally, for $\Rey=4000$, the thickness $\delta$ is estimated in the same manner. As shown in \fref{fig:delta}(b), the spectra in the vicinity of the wall have the energy peak at $\lambda_\theta= \pi$, which corresponds to the observation of mode-2 and, as highlighted in blue and red, the peak gradually switches from $\lambda_\theta= \pi$ to a shorter one with $y_\mathrm{out}^+=15.8 \to 30.1$. 
As a result, we determine $\delta^+_\mathrm{out}=0.31$ for $\Rey=4000$. 

The properties of every modal structure found on the outer cylinder side of $\eta=0.2$ are summarised in \tref{tab:mode}, including the thickness of the structure that was estimated as described above. 
Note that the thickness $\delta_{\rm out}$ could not be determined in the case of $\Rey=500$, because the peak shift relevant to the structure variation from the outer to the inner side was unclear at such a low Reynolds number.
According to this result, the critical Reynolds number of the transition from/to mode-2 exists between $\Rey=1250$ and 1375. 
The thickness $\delta_{\rm out}^+$ increases monotonically with the increase of the Reynolds number. 
However, the aspect ratio $\lambda/\delta$ appears to remain almost constant, particularly for the mode-2 structure.
It may be observed that $\lambda/\delta$ decreases gradually as $\Rey$ increases up to 2000, but starts to increase when $\Rey=2000 \to 4000$. 
This change in the Reynolds number dependency is consistent with the tendency of $u'_{\theta, {\rm rms}}$ in \fref{fig:Rewrms}, where only $\Rey=4000$ provides a qualitatively different profile in comparison to the smaller values. 
This observation implies that there might be a further transition(s) in the flow structure on the outer cylinder side at higher Reynolds numbers. 
An investigation using a large-scale DNS will be considered for future work. 

\begin{table}
\caption{Properties of modal structure found on outer cylinder side for $\eta=0.2$ at different Reynolds numbers. Here, $\lambda_\theta$ is the wavelength in the azimuthal direction (in rad), $\delta_{\rm out}^+=\delta u_{\tau,{\rm out}}/\nu$ is the radial thickness of the structure scaled by the wall units, and $\lambda/\delta = 2\Rey_{\tau,{\rm out}} \lambda_\theta/(1-\eta)\delta_{\rm out}^+$ is the aspect ratio of the modal structure.}
\label{tab:mode}
\begin{tabular}{llllllll}
\hline\noalign{\smallskip}
$\Rey$              &  500 & 1000 & 1250 & 1375 & 1500 & 2000 & 4000 \\
\noalign{\smallskip}\hline\noalign{\smallskip}
Mode                & 1    & 1    & 1    & 2    & 2    & 2    & 2    \\
$\lambda_\theta$    &$2\pi$&$2\pi$&$2\pi$&$\pi$ &$\pi$ &$\pi$ &$\pi$ \\
$\delta^+_\mathrm{out}$& ---  & 13.5 & 13.3 & 13.6 & 15.8 & 20.4 & 30.1 \\
$\lambda/\delta$    & ---  & 16.5 & 19.7 & 10.3 & 9.5  &  9.2 & 11.0 \\
\noalign{\smallskip}\hline
\end{tabular}
\end{table}

\section{Conclusion}

In this numerical study, we investigated the flow structure and turbulent statistics of the counter-rotating Taylor--Couette flows with relatively small radius ratios of $\eta = 0.2$--0.5, over a wide range of the Reynolds number, from the laminar to the turbulent regime. 
On the inner (unstable) side, the Taylor-vortex caused by the centrifugal instability dominated the flow field, and this tendency appeared to be independent of the radius ratio. 
The flow field on the outer (stable) side was rather passive against the inner side Taylor-vortex flow.
However, it was shown that there existed a critical value of the radius ratio between $\eta=0.2$ and 0.3, below which the Taylor-vortex on the inner-cylinder side did not exert a significant influence on the outer side flow. 
For all of the tested Reynolds numbers with a radius ratio as small as $\eta=0.2$, one common pattern of particular interest was the modal structure that was homogeneous in the axial direction, periodic in the azimuthal direction, and resembled the TS-instability wave. 
The modal structure found on the outer cylinder (i.e., stable) side in the current flow configuration may had been triggered by similar mechanisms as the TS wave in the rotating plane channel flow.  
The modal structure changed the azimuthal wavenumber at a certain value between $\Rey=1250$ and 1375. 
The thickness of the structure increased as the Reynolds number increased, but the wavelength to thickness ratio of the mode-2 structure remained almost constant. 
The modal structure did not contribute to the Reynolds shear stress (i.e., the radial momentum transfer), although the magnitude of the azimuthal velocity variation was comparable to that of the inner side turbulent and Taylor-vortex flows.

The current Reynolds-number range is still limited to early-stage turbulence. Hence, further DNS studies with higher Reynolds numbers are required. 
Moreover, the various Taylor--Couette flows with different wall velocities between two cylinders remain open issues to be addressed in future work.
Through additional studies, a more detailed mechanism of the modal structure could be elucidated. 
As a first step, the finding of this study with regard to the coexistence of the modal structures and the Taylor-vortices in a closed flow configuration may clarify the instabilities in subcritical flow.

\begin{acknowledgements}
T.K.~was received support from the Japan Society for the Promotion of Science (JSPS), Fellowship \#17J04115. 
T.T. received support from the JSPS KAKENHI Grants, \#16H06066.
This study was partly carried out with the large-scale computer systems at the Cyberscience Centre, Tohoku University, and the systems at the Cybermedia Centre, Osaka University.
\end{acknowledgements}

%

\end{document}